\documentclass[twocolumn]{article}

\usepackage[twocolumn,textwidth=18cm,columnsep=.81cm]{geometry}
\usepackage{amsmath}
\usepackage{amssymb}
\usepackage{amsthm}
\usepackage{physics}
\usepackage{graphicx}
\usepackage{hyperref}
\usepackage{dcolumn}
\usepackage{stfloats}
\usepackage{authblk}
\usepackage{caption}
\usepackage{quantikz}
\usepackage{needspace}
\usepackage{helvet}
\usepackage{tikz, pgfplots}
\usepackage{mathtools}
\pgfplotsset{compat=1.18}
\usepackage{xcolor}
\usepackage{appendix}
\usepackage[T1]{fontenc}
\usepackage{booktabs}
\usepackage{array}
\usepackage{subcaption}
\usepackage{indentfirst}
\usepackage[numbers, sort&compress]{natbib}
\title{The Cross-Kernel Margin: A Robustness Measure for Quantum Kernel Methods}

\author[1,2]{S. Govender}
\author[1,2,*]{I. Sinayskiy}
\small \affil[1]{Discipline of Physics, School of Agriculture and Science, University of KwaZulu-Natal, Durban 4001, South Africa}

\small \affil[2]{National Institute for Theoretical and Computational Sciences (NITheCS), KZN Node, University of KwaZulu-Natal, Durban 4001, South Africa.}

\small \affil[*]{Corresponding Author: sinayskiy@ukzn.ac.za}

\date{}

\begin{document}
\twocolumn[
  \begin{@twocolumnfalse}
  \maketitle
\vspace{-2.5em}
    \begin{abstract}
\noindent Quantum devices in the current Noisy Intermediate-Scale Quantum (NISQ) era are inherently affected by noise, which can degrade the predictive performance of quantum machine learning models. In this work, we introduce a new margin-based robustness measure for Quantum Kernel-Assisted Support Vector Machines (QSVMs), termed the \textit{cross-kernel margin}. This measure quantifies the stability of a classifier learned under a perturbed kernel relative to the ideal feature space. We derive \textit{a posteriori} stability bounds for the corresponding cross-kernel inverse squared-margin under kernel perturbations using the Tikhonov-stabilised SVM dual formulation. The local depolarising noise model is then applied to this framework to induce perturbations in the kernel. The resulting bounds are numerically checked using simulations across multiple datasets and further tested using kernel matrices obtained from real quantum hardware and a noisy backend simulator. Furthermore, we empirically compare the degradation of test accuracy under local depolarising noise with the commonly used global depolarising noise model in order to motivate its use in our study. Finally, we present
empirical results linking margin-based quantities with the
generalisation performance of QSVMs, providing additional motivation for
our margin-based robustness analysis. 
   \end{abstract}
  \end{@twocolumnfalse}
]

\section{Introduction}
\vspace{-0.5em}

The field of quantum machine learning (QML) lies at the intersection of quantum computing and
classical machine learning (CML) \cite{schuld2015introduction}. However, practical applications of QML are limited by the unavoidable presence of noise prevalent in the current Noisy Intermediate-Scale Quantum (NISQ) era \cite{preskill2018quantum}. Noise results in errors which degrade model performance and limit the practical implementation of QML models. 

In the context of model performance, robustness broadly refers to the ability of a machine learning model to maintain its predictive performance when subjected to perturbations, environmental changes or shifts in the underlying data distribution used to train a model \cite{freiesleben2023beyond}. Such perturbations may result from corrupted data, adversarial perturbations or measurement errors \cite{zhu2004class}. Since device noise associated with quantum hardware can affect the feature representation of data used by a QML model, which can degrade its performance, it is particularly important to understand the robustness of QML models to noise in the NISQ era 
\cite{zhu2004class, padro2026robustness,jin2026noise, bose2026systematic}.

Generalisation and robustness have been shown to be distinct but closely related concepts \cite{kawaguchi2022robustness}. 
Generalisation describes the ability of a machine learning model to learn patterns from training data and successfully apply those patterns to previously unseen samples drawn from the same underlying data distribution. \citet{xu2012robustness} have demonstrated a theoretical connection between generalisation and robustness by showing that a learning algorithm must be robust to some degree in order to generalise effectively. This connection has also been shown via data-dependent analyses \cite{kawaguchi2022robustness}. This link suggests that quantities that successfully characterise the generalisation behaviour of a classifier may also provide insight into its robustness.

Generalisation has been extensively studied in the field of CML and has been quantified using several measures \cite{langford2002pac, herbrich2000pac,mcallester2003simplified, mohri2018foundationsML,gronlund2020near}. In particular, a strong theory of generalisation has been developed for classical support vector machines (SVMs) \cite{bartlett1999generalization,vapnik1968uniform, vapnik2000bounds, vapnik2006estimation,langford2002pac,herbrich2000pac, gronlund2020near}, 
since their introduction by \citet{boser1992training} and subsequent extension to non-separable datasets by \citet{cortes1995support}.

In contrast, generalisation in QML remains relatively underdeveloped. Recent work involving near-term QML models has revealed unexpected generalisation behaviour, which has challenged the applicability of traditional uniform generalisation bounds to near-term models \cite{caro2022generalization, gil2024understanding}. These findings suggest that an alternative, data-dependent approach may be necessary to study the generalisation behaviour of modern QML models.

Margin-based metrics have since emerged as a promising alternative to studying generalisation. Margin-based generalisation measures have recently been shown to be stronger predictors of generalisation performance than conventional metrics for Quantum Neural Networks (QNNs) \cite{hur2024understanding}. However, analogous margin-based quantities remain largely unexplored for Quantum-Kernel-Assisted Support Vector Machines (QSVMs). Given the connection between robustness and generalisation, margin-based quantities may also provide useful insights for studying the robustness of QSVMs under noise.

A common approach to modelling quantum hardware noise is by using the depolarising noise model. Under maximum depolarising noise, a quantum state reduces to the maximally mixed state, effectively losing all useful information required for the learning task. Consequently, understanding the impact of noise on the predictive power of modern QML models has become an active area of research \cite{wang2021towards, wang2025power, khanal2024generalization}. Existing studies, however, primarily focus on the global depolarising model, in which all qubits in the system are affected by noise simultaneously. While analytically tractable, this model does not accurately reflect real device noise. The local depolarising noise model, which applies noise independently to each qubit, provides a more realistic description of device noise but remains relatively unexplored \cite{nielsen2010quantum}. 

In this work, we introduce the \textit{cross-kernel margin} as a margin-based robustness measure for QSVMs trained using perturbed kernel matrices. This quantity evaluates the dual solution learned from a perturbed kernel with respect to the geometry of the ideal kernel. This quantifies the stability of the classifier learned in the presence of a perturbation relative to the ideal feature space. We derive analytical stability bounds for the corresponding inverse squared cross-kernel margin under kernel perturbations using the Tikhonov-stabilised SVM dual formulation \cite{tikhonov1943stability}. The local depolarising noise model is then used as an application to induce perturbations in the quantum kernel. The resulting bounds are numerically validated using simulations across multiple datasets, and further tested using quantum kernels obtained from experiments on real quantum hardware and a noisy backend simulator for multiple datasets. 

We further motivate our use of local depolarising noise as the primary application of our stability bounds through numerical simulations comparing QSVM model performance under both local depolarising noise and the widely-used global depolarising noise. Additionally, to motivate the use of margin-based robustness measures in this study, we first provide empirical evidence of an association between an analogous \textit{cross-label margin} and the generalisation performance of QSVMs in the ideal setting. Altogether, these results provide insight into the robustness of QSVMs to noise-induced kernel perturbations in the NISQ era. 

Throughout this work, we distinguish between robustness as a property of the classifier under perturbations and stability as the mathematical characterisation of how much a quantity changes under such perturbations. The cross-kernel margin is introduced in this study as a robustness measure, while the analytical results are stability bounds describing the sensitivity of this robustness measure to kernel perturbations. 

The paper is structured as follows. Sections \ref{sec: LitReview} and \ref{sec: Theory}  present a literature review and provide the necessary background on margin theory for classical and quantum-enhanced SVMs. Section \ref{sec: MarginGenLink} presents empirical evidence  linking margin-based quantities to generalisation performance in the ideal setting. Section \ref{sec: LocalVGlobal} compares QSVM model performance under local and global depolarising noise. The main analytical results of the paper are presented in Section \ref{sec: Results}, while numerical and hardware experiments testing these results are presented in Section \ref{sec: NumericalExp}. Finally, Section \ref{sec: Conclusion} concludes the paper.

\section{Literature Review}\label{sec: LitReview}

Robustness to noise is an important property of any machine learning (ML) model. Classically, noise is viewed as a form of data distortion and may be categorised as either attribute noise or class noise \cite{zhu2004class}. Attribute noise refers to corrupted instances of the dataset associated with training, such as missing, inaccurate or erroneous measurements, while class noise occurs when training examples have been assigned incorrect class labels \cite{zhu2004class}. Noise present in input data has been shown to degrade predictive performance of classical models \cite{zhu2004class, saez2016evaluating}. Similarly, feature noise and class imbalance negatively affect the performance of QML models \cite{sheoran2025robust, jin2026noise, bose2026systematic}. 

The notion of robustness can be studied from several perspectives, such as adversarial robustness \cite{biggio2011support, kustiawan2026adversarial}, algorithmic robustness \cite{xu2012robustness} and robustness to shifts in data distribution or predictive performance \cite{freiesleben2023beyond}. In this work, robustness is considered in terms of the stability of the learned classifier under noise-induced perturbations. 

Traditional metrics used to measure robustness in ML models include accuracy, precision, recall, and derived measures such as the equalised loss of accuracy \cite{ferri2009experimental, ishii2021comparative, saez2016evaluating}. However, recent work has suggested that these task performance measures may not be entirely informative in terms of prediction stability under distribution shifts  \cite{carvalho2025rethinking}.  Alternatively, agreement-based robustness measures have been proposed that are able to quantify the consistency between predictions based on data obtained from clean and perturbed data distributions \cite{padro2026robustness}. 

While robustness and generalisation are distinct concepts \cite{freiesleben2023beyond}, studies have established a connection between them \cite{xu2009robustness}. \citet{xu2009robustness} established that a robust learning algorithm caused regularised SVMs to generalise successfully, while \citet{kawaguchi2022robustness} linked the two concepts via a data-dependent analysis. Due to this link, it appears that successful measures established for generalisation may also provide insight into robustness.  

Generalisation theory for classical Support Vector Machines (SVMs) is relatively well-developed. Following the introduction of SVMs by \citet{boser1992training} and the subsequent extension to non-separable datasets by \citet{cortes1995support}, foundational work by \citet{bartlett1999generalization} established generalisation guarantees for SVMs using margins. Following this, subsequent studies derived generalisation bounds based on Vapnik-Chervonenkis (VC) theory \cite{vapnik1968uniform}, the expected number of support vectors \cite{vapnik2000bounds}, Rademacher complexity \cite{bartlett2002rademacher} and PAC-Bayesian margin bounds \cite{herbrich2000pac, langford2002pac, mcallester2003simplified}. More recently, \citet{gronlund2020near} derived near-tight upper and lower bounds for SVM generalisation in terms of margins. 

Several analogous approaches have been proposed for studying generalisation in QML. These include analyses based on pseudo-dimension \cite{caro2020pseudo}, the quantum Fisher information metric \cite{banchi2021generalization, quantumFisher2024}, and Rademacher complexity in the context of parametrised quantum circuits \cite{bu2022statistical, bu2023effects, bu2021rademacher}, with \citet{bu2021rademacher} exploring the effects of noise in the quantum circuit. A comprehensive review of existing generalisation measures and error bounds for QML models in the NISQ era can be found in \cite{khanal2024generalization}.

Despite these developments, recent findings have revealed unexpected generalisation behaviour by near-term quantum devices. Systematic randomisation experiments have highlighted the ability of modern deep learning models to memorise randomly labelled datasets \cite{zhang2021understanding}. Similar results were later reported by \citet{gil2024understanding}, who performed similar experiments for near-term QNNs. Furthermore, \citet{caro2022generalization} observed strong generalisation performance in QML models trained on relatively small datasets, challenging conventional assumptions regarding training sample sizes. These findings have triggered an increasing interest in alternative dataset-dependent measures to indicate generalisation performance instead. Margin-based metrics have shown particular promise in this regard \cite{bartlett2017spectrally, hur2024understanding}. 

Several studies have investigated the impact of noise on the predictive power of QML models \cite{wang2021towards,thanasilp2024exponential, wang2025power}. In particular, \citet{thanasilp2024exponential} demonstrated that under specific conditions, quantum kernel values can become exponentially concentrated around a fixed value, potentially producing trivial models with poor predictive performance. These studies provide vital insight into robustness and generalisation in the presence of noise. However, to the best of our knowledge, no margin-based robustness measures have been proposed for QSVMs under noise.

This gap motivates the introduction of the cross-kernel margin proposed in this work, which quantifies the stability of the learned classifier under kernel perturbations, such as those induced by noise. This development thus provides a margin-based framework for analysing the robustness of QSVMs in the NISQ era.

\section{Preliminaries}
\label{sec: Theory}
\subsection{SVM Theory}\label{sec: SVMTheory}
In the simplest case, to perform binary classification on linearly separable data using a linear classifier, the SVM algorithm attempts to find a hyperplane (which is a line in two dimensions) that effectively separates data points belonging to two different classes. For separable data, infinitely many hyperplanes exist that could satisfy this criterion; hence, by solving an optimisation problem, the SVM algorithm locates the optimal separating hyperplane such that the distance from the hyperplane to the nearest data points from each class is maximised. 

This distance is termed the geometric margin, which is defined as the minimum Euclidean distance from all points to the hyperplane. The closest data points to the hyperplane are referred to as support vectors.  

For the linearly separable case, in two dimensions, this hyperplane is a line, with the equation $\boldsymbol{w} \cdot \boldsymbol{x} + b = 0$. Here, $\boldsymbol{w}$ is termed the weight vector, and is perpendicular (normal) to the hyperplane. $b$ is the bias term (or intercept), which shifts the hyperplane away from the origin. 

The functional (confidence) margin of a classifier $f$ at point $\boldsymbol{x}_i$ is defined as $y_if(\boldsymbol{x}_i)$ where $y_i \in \{-1,+1\}$ is the true label of the point $\boldsymbol{x}_i$ \cite{mohri2018foundationsML}. For binary classification problems, the SVM classifier returns the hypothesis $f\xrightarrow[]{} \text{sign}(\boldsymbol{w} \cdot \boldsymbol{x} + b)$, such that all points of one class are given a positive value and are placed on one side of the hyperplane, whereas the points belonging to the other class hold a negative value and fall on the other side of the hyperplane. In this way, it is clear that when $y_if(\boldsymbol{x}_i) > 0$, the sample has been correctly classified. 

For real-valued functions, $f$, functional margins can be calculated using hypotheses returned as $f\xrightarrow[]{}\boldsymbol{w} \cdot \boldsymbol{x} + b$. In this case, the magnitude $|f(\boldsymbol{x}_i)|$ represents the \textit{confidence} of the prediction made by the classifier \cite{mohri2018foundationsML}. In the linearly separable case, the functional and geometric margins are related for each training sample $\boldsymbol{x}_i$ as, $|y_if(\boldsymbol{x}_i)|\geq \gamma \|\boldsymbol{w}\|$, where $\|\cdot \|$ represents the Euclidean (L2) norm and $\gamma$ is the geometric margin.

Through rescaling of the real-valued classifier output, detailed in \cite{mohri2018foundationsML}, the maximum geometric margin takes the form,

\begin{equation}
\gamma = \max_{(\boldsymbol{w}, b)} \frac{1}{\|\boldsymbol{w}\|}
\end{equation}

\noindent where the maximisation is over all separating hyperplanes satisfying $y_i(\boldsymbol{w} \cdot \boldsymbol{x}_i + b) \geq 1$.

\begin{figure}
    \centering
    \includegraphics[width=1\linewidth, height = 6cm]{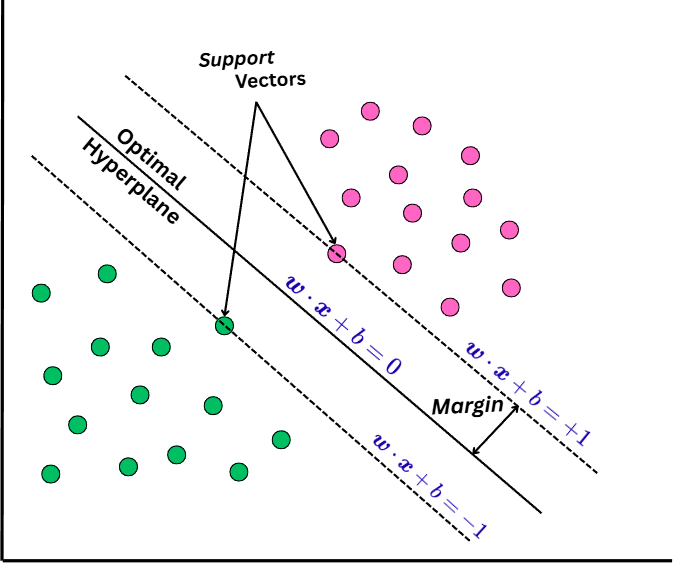}
    \caption{Diagram depicting the optimal linear hyperplane returned by an SVM to separate two classes of data points (green and pink circles). The support vectors used to determine the hyperplane and the corresponding geometric margin are labelled. The equations of the separating and marginal hyperplanes are depicted in blue. A hard margin is depicted here since all samples are classified with a distance greater than the geometric margin away from the separating hyperplane.}
    \label{fig:SVM_HardMargin}
\end{figure}

Figure 
\ref{fig:SVM_HardMargin} depicts the optimal hyperplane that separates two distinct classes (green and pink circles), and in doing so, obtains the maximal margin. The marginal hyperplanes, with equations $\boldsymbol{w} \cdot \boldsymbol{x} + b = \pm 1$ are shown parallel to the separating hyperplane and the support vectors used to determine the margin are portrayed accordingly.

\begin{figure}
    \centering
    \includegraphics[width=1\linewidth, height = 7cm]{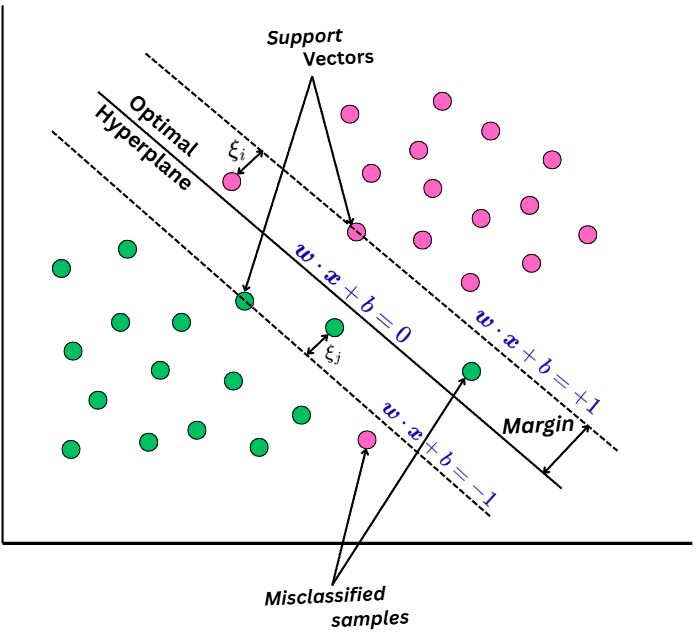}
    \caption{Diagram depicting the optimal linear hyperplane returned by the SVM to separate the two classes of data points (green and pink circles). The equations of the separating and marginal hyperplanes are depicted in blue. A soft margin is depicted here since some points lie within the margin region but have corresponding positive slack variables $\xi_i$ and $\xi_j$. Misclassified points which are placed on the wrong side of the hyperplane are also labelled.}
    \label{fig:SVM_SoftMargin}
\end{figure}

In the case where the samples corresponding to each class cannot be linearly separated, then the previously discussed constraints may not hold. This case is illustrated in Figure \ref{fig:SVM_SoftMargin}, which depicts misclassified samples and samples from each class correctly classified, but with a Euclidean distance that is less than the maximum margin distance from the hyperplane.

There are two formalisms which describe the optimisation problem defining the SVM algorithm; namely the primal and dual formalisms.

The primal optimisation problem for the non-separable case is defined by the objective function,
\begin{equation}\label{primalOpt}
    \min_{\boldsymbol{w}, b, \boldsymbol{\xi}} \frac{1}{2} \|\boldsymbol{w}\|^2 + C\sum_{i=1}^m \xi_i
\end{equation}
\noindent which satisfies the constraints

\begin{equation}\label{primalConstraint}
y_i (\boldsymbol{w} \cdot \phi(\boldsymbol{x}_i)+b) \geq 1-\xi_i 
\text{ and } 
  \xi_i  \geq 0 \text{ for } i \in [1, m] 
\end{equation}

\noindent Here, $\boldsymbol{\xi} = (\xi_1 \dots \xi_m)^T$ denotes the vector of slack variables, where $\xi_i$ measures the extent to which the margin constraint is violated by sample $\boldsymbol{x}_i$. Samples with $0<\xi_i<1$ lie inside the margin but remain correctly classified, whereas samples with $\xi_i>1$ are misclassified. The parameter $C$ controls the trade-off between maximising the margin and penalising margin violations. Larger $C$ values impose stricter penalties on violations of the margin constraints, while smaller values allow larger violations in favour of maximising the separating margin. Here, $m$ denotes the number of labelled samples in a training dataset and $\phi(\cdot)$ represents a possible feature map, which is a notion that is introduced at a later stage.

The maximal margin obtained in this case is termed the soft margin, since it allows violations of the margin constraints through the introduction of slack variables. By contrast, the margin for the linearly separable case is termed the hard margin and does not permit any margin violations.

The Lagrangian associated with the primal formalism \cite{mohri2018foundationsML} is defined as 
\begin{equation}
\begin{split}
     \mathcal{L} ( \boldsymbol{w}, b, \boldsymbol{\xi}, \boldsymbol{\alpha}, \boldsymbol{\beta}) = \frac{1}{2} \|\boldsymbol{w}\|^2 + C\sum_{i=1}^m \xi_i -
    \sum_{i=1}^m \\ \alpha_i[y_i(\boldsymbol{w} \cdot \phi(\boldsymbol{x}_i)+b)-1+\xi_i] - \sum_{i=1}^m \beta_i \xi_i
    \end{split}
\end{equation}

\noindent with Lagrangian variables $\boldsymbol{\alpha} = (\alpha_1, ..., \alpha_m)^T$ and $\boldsymbol{\beta} = (\beta_1, ..., \beta_m)^T$,where  $\alpha_i \geq 0$ and $\beta_i \geq 0$, $i\in [1,m]$ are Lagrange multipliers associated with the constraints defined in expression (\ref{primalConstraint}).

The primal Lagrangian can be used to obtain the Karush-Kuhn-Tucker (KKT) conditions for the non-separable case as,
\begin{align}
    \boldsymbol{w} &= \sum_{i=1}^m \alpha_i y_i \phi(\boldsymbol{x}_i)   \\
    0 &= \sum_{i=1}^m \alpha_i y_i \\
    C &= \alpha_i + \beta_i, \text{ for all } i\in \{1, \cdots ,m\}\\
    \alpha_i &= 0 \, \text{ or } \, y_i(\boldsymbol{w} \cdot \phi(\boldsymbol{x}_i) + b) = 1 - \xi_i\\
    \beta_i &=0 \, \text{ or } \, \xi_i = 0 
\end{align}

\noindent The complete derivation of these conditions can be found in \cite{mohri2018foundationsML}.

It follows from the penultimate condition that whenever $\alpha_i>0$, the corresponding sample satisfies 
$$
y_i(\boldsymbol{w} \cdot \phi(\boldsymbol{x}_i) + b) = 1 - \xi_i$$
\noindent and therefore corresponds to a support vector. Additionally, if $0 < \alpha_i < C$, then $\beta_i>0$ implying that $\xi_i = 0$, then these corresponding support vectors $\boldsymbol{x}_i$ lie on the margin. 

The Lagrangian for the dual formalism can then be obtained by substituting the first KKT condition into the primal Lagrangian and then applying the second condition. This Lagrangian has the form, 

\begin{equation}\label{lagrangian}
    \mathcal{L} ( \boldsymbol{\alpha}) = 
    \sum_{i=1}^m \alpha_i  - \frac{1}{2} \sum_{i,j=1}^m \alpha_i \alpha_j y_i y_j ( \phi(\boldsymbol{x}_i) \cdot \phi(\boldsymbol{x}_j))
\end{equation}

The SVM optimisation problem is solved by minimising the primal Lagrangian with respect to the primal variables ($\boldsymbol{w}, b, \boldsymbol{\xi}$) and maximising the dual Lagrangian with respect to the dual variables ($\boldsymbol{\alpha}$), subject to the constraints 
$$
0 \leq \alpha_i \leq C \text{ and } \sum_i \alpha_i y_i = 0
$$  
\noindent for $i \in [1,m]$ for $m$ training samples.
The dual Lagrangian function has been largely studied since it can be easily modified to accommodate the famous kernel trick \cite{boser1992training}
by replacing the dot product of the support vectors with the kernel element associated with them in the SVM algorithm. This enables the algorithm to utilise a feature map to map the data to a higher-dimensional space (feature space) in which a hyperplane (higher-dimensional surface) can be found in order to separate the data. The kernel trick is beneficial since there is no need to explicitly define the feature map, $\phi(\cdot)$. It is only necessary to determine the inner product in this space. 

Modifying equation (\ref{lagrangian}) to include the kernel results in the form
\begin{equation}
  \mathcal{L} (\boldsymbol{\alpha}, K) = \sum_{i=1}^m \alpha_i - \frac{1}{2}\sum_{i,j=1}^m \alpha_i \alpha_j y_i y_j K( \boldsymbol{x}_i, \boldsymbol{x}_j)
\end{equation}

\noindent where $K(\boldsymbol{x}_i, \boldsymbol{x}_j)$ computes the kernel value for samples $\boldsymbol{x}_i$ and $\boldsymbol{x}_j$. Additionally, we can define the quantity, 
\begin{equation}\label{sqNormweight}
\|\boldsymbol{w}\|^2 =\sum_{i,j=1}^m \alpha_i \alpha_j y_i y_j K( \boldsymbol{x}_i, \boldsymbol{x}_j)= \boldsymbol{{\alpha}}^TYKY\boldsymbol{\alpha}
\end{equation}
\noindent where the notation $\|\boldsymbol{w}\|^2$ denotes the squared Reproducing Kernel Hilbert Space (RKHS) norm of the weight vector associated with the feasible dual solution $\boldsymbol{\alpha}$ \cite{mohri2018foundationsML}. The inverse of this quantity yields the squared geometric margin of the form,

\begin{equation}\label{marginDef}
\gamma^2 = \frac{1}{\|\boldsymbol{w}\|^2} 
\end{equation}

\subsection{QSVM Theory}

Notably, the kernel trick, and by extension, kernel methods are essential for QSVMs. QSVMs use quantum circuits run on a quantum computer to obtain a matrix of kernel values \cite{schuld2019quantum}. These matrix elements are commonly computed as the squared state overlap between two feature vectors corresponding to two data points. These kernel elements take the form,
\begin{equation}
K(\boldsymbol{x}_i, \boldsymbol{x}_j) = |\langle \phi(\boldsymbol{x}_j )|\phi(\boldsymbol{x}_i)\rangle |^2
\end{equation}

\noindent where $|\phi(\boldsymbol{x}_i)\rangle$ defines a feature map, which maps the classical data point $\boldsymbol{x}_i$ to a higher-dimensional Hilbert space.

The Hilbert-Schmidt inner product \cite{nielsen2010quantum} can be used to calculate the kernel elements for mixed states as, $K(\boldsymbol{x}_i, \boldsymbol{x}_j) = \mathrm{Tr}(\rho(\boldsymbol{x}_i) \rho(\boldsymbol{x}_j))$ where $\rho(\boldsymbol{x}_i)$ represents the density matrix associated with the data point $\boldsymbol{x}_i$, which reduces to the squared state overlap expression for pure states.
 Notably, the computation of the matrix of the quantum kernel values is the only process run on the quantum computer. The quantum kernel (Gram) matrix is then used by the SVM algorithm in a manner identical to the classical formalism. Thus, the resulting QSVM algorithm is denoted as a hybrid quantum-classical ML algorithm.  

\subsection{Depolarising Noise Models}\label{sec: DepolNoiseModels}
A depolarising noise channel is a
completely positive trace-preserving (CPTP) map that maps a quantum state onto a maximally mixed
state with some finite probability \cite{nielsen2010quantum}.

Global depolarising noise channels are applied to the entire system of qubits simultaneously, whereas local depolarising noise channels act independently on each qubit
in the system. Despite being analytically tractable, global depolarising noise does not realistically model noise on near-term hardware, where errors occur at the qubit level. The resulting $N$-qubit state after the application of a global depolarising channel is given by the following expression,

\begin{equation}\label{globalDepol}
    \mathcal{E}_{G}(\rho)=(1-p_G)\rho + p_G\frac{I}{2^N}
\end{equation}

\noindent where $p_G$ refers to the probability that the global depolarising channel replaces the state $\rho$ with the maximally-mixed state, and $I$ is the identity matrix.

The single-qubit state after the application of a local depolarising channel can be expressed using the Pauli matrices, $(\sigma_x, \sigma_y, \sigma_z)$ as \cite{nielsen2010quantum}

\begin{equation}
    \mathcal{E}_L(\rho) = (1-\frac{3}{4}p'_L)\rho + \frac{p'_L}{4}(\sigma_{x}\rho \sigma_{x} + \sigma_{y} \rho \sigma_{y} + \sigma_{z} \rho \sigma_{z})
\end{equation}

\noindent where $p'_L$ is the local depolarising noise probability. Since $p_G$ and $p_L'$ represent depolarising noise probabilities, it is clear that they must have a value in the range $[0,1]$.

The above expression can be rewritten by defining the scaled parameter $p_L = \frac{3}{4} p_L'$, as,

\begin{equation}\label{localDep}
    \mathcal{E}_L(\rho) = (1-p_L)\rho + \frac{p_L}{3}(\sigma_{x}\rho \sigma_{x} + \sigma_{y} \rho \sigma_{y} + \sigma_{z} \rho \sigma_{z})
\end{equation}

\noindent where $p_L \in [0, \frac{3}{4}]$.

The Kraus (operator-sum) representation can also be used to describe the single-qubit local depolarising channel acting on some state $\rho$, given by $\mathcal{E}_L(\rho)$, 

\begin{equation}\label{KrausSum}
    \mathcal{E}_L(\rho) = \sum_{i=0}^3 K_i \rho K_i^{\dagger}
\end{equation}

\noindent where for some probability $p_L \in [0, \frac{3}{4}]$, the Kraus operators for a single-qubit, are defined as,

\begin{align}\label{krausOp}
    K_0 = \sqrt{1-p_L}\sigma_0
\end{align}
\begin{align}
    K_1 = \sqrt{\frac{p_L}{3}}\sigma_x,\;
    K_2 = \sqrt{\frac{p_L}{3}}\sigma_y, \;
    K_3 = \sqrt{\frac{p_L}{3}}\sigma_z
\end{align}

The local depolarising noise channel may be interpreted as applying a random Pauli error to the original state with some probability, $p_L$. For the noiseless case, the channel reduces to the identity channel. At $p_L=\frac{3}{4}$, the channel becomes fully depolarising. The application of a fully depolarising channel maps every input state to the maximally mixed state. 

\section{Linking Margins to Generalisation}\label{sec: MarginGenLink}

In this section, we investigate the link between the cross-label margins and QSVM generalisation performance in the ideal setting. Taking inspiration from the systematic randomisation experiments performed in \cite{gil2024understanding} and \cite{hur2024understanding}, we explore the effect of label corruption on test accuracy, which is used here as our proxy for generalisation performance. Additionally, we investigate the effect of corrupting increasing fractions of the labels in the training dataset on the margin distribution across multiple datasets.

Since we are in the noiseless regime here, we use an analogous metric to the cross-kernel margin, namely the \textit{cross-label margin}. Analogously to the cross-kernel margin, this quantity is defined using the ideal kernel matrix with the noisy dual solution. However, in this case, the noisy dual solution arises from corrupted ("noisy") training labels, not noise affecting the kernel elements. As such, the ideal kernel is used throughout the analysis and the cross-comparison in this case is with regard to the training label distribution. 

The margin distribution is studied by computing the per-sample cross-label margin, defined analogously to the geometric margin. For a training sample, $\boldsymbol{x}_i$, this quantity is given by 
${y_i \tilde{f}(\boldsymbol{x}_i)}/{\|\boldsymbol{w}_{\boldsymbol{\tilde{\alpha}}, \tilde{y}}\|}$, 
where $\tilde{f}(\boldsymbol{x}_i)$ denotes the decision function obtained from a QSVM trained using corrupted labels, and $\|\boldsymbol{w}_{\boldsymbol{\tilde{\alpha}}, \tilde{y}}\|^2 = \sum_{i,j=1}^m \tilde{\alpha}_i \tilde{\alpha}_j \tilde{y}_i \tilde{y}_j {K}_{ij}$ is the corresponding squared RKHS norm evaluated using the corrupted training labels. Explicitly this decision function takes the form,  $\sum _{j=1}^m \tilde{\alpha}_j\tilde{y}_j K(\boldsymbol{x}_j, \boldsymbol{x}_i) + \tilde{b}$
where ${\tilde{\alpha}_j}$,  $\tilde{y}_j$ and $\tilde{b}$ denote the dual variables,  corrupted training labels and the corresponding corrupted bias term respectively. Notably, the clean labels $y_i$ are used to compute the cross-label margin to ensure that the resulting measure quantifies the agreement of the classifier learned from the corrupted labels and the original (ideal) classification task. 

Using cross-validation on various datasets, we analyse the relationship between the median cross-label margin and test accuracy under controlled corruption of the training labels in each dataset. This relationship was explored using four datasets, namely the High Time Resolution Universe Survey (South) (HTRU2) astrophysical dataset \cite{lyon2016fifty}, a synthetic dataset describing a Gaussian distribution and the (White) Wine Quality and Heart Disease datasets. These datasets are further described in Appendix \ref{AppendixE}. 

\begin{figure*}[t]
    \centering
    \includegraphics[width=1\linewidth, height = 5cm]{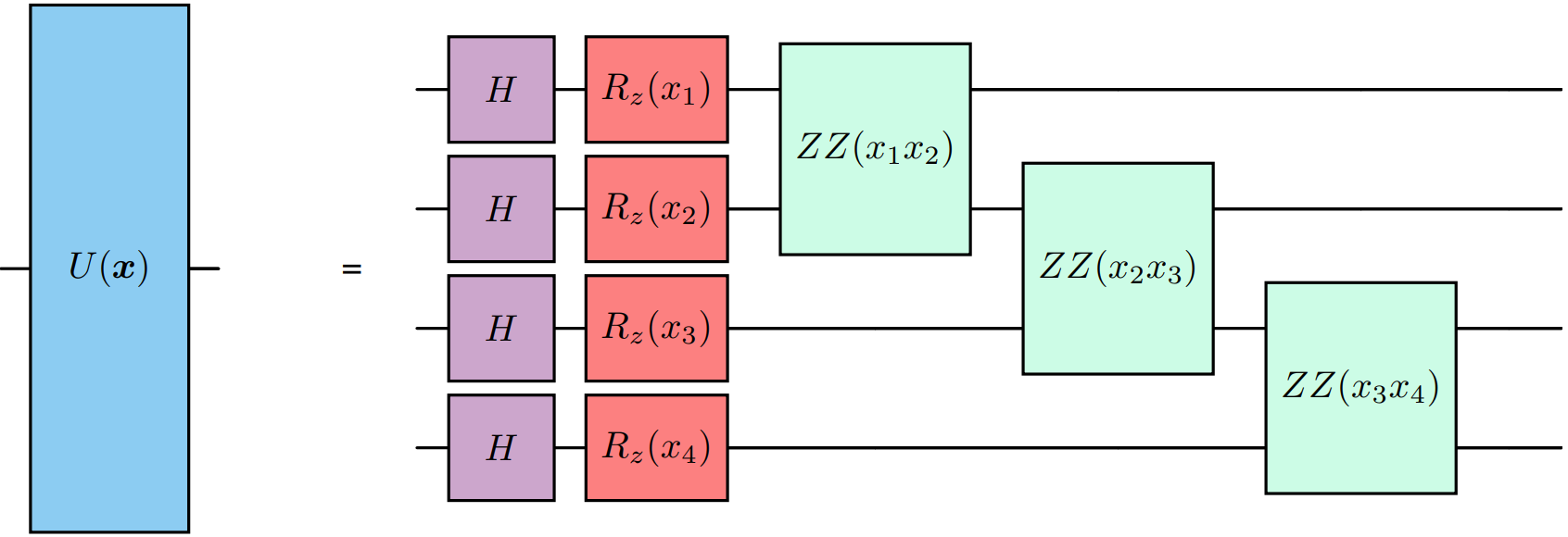}
    \caption{Single layer of the Instantaneous Quantum Polynomial-time (IQP) encoding with nearest-neighbour entanglement for four qubits. This quantum circuit uses Hadamard gates (purple), parametrised $R_z$ rotations about the $z$-axis (red), and ZZ entangling gates (green), where the input features $\boldsymbol{x}_i$ determine the rotation angles.}
    \label{fig:IQP}
\end{figure*}

Each dataset was encoded using Instantaneous Quantum Polynomial-time (IQP) embedding \cite{havlivcek2019supervised}, following its previous use in literature \cite{wang2025power}. This was implemented using a quantum circuit consisting of two qubits with a single application of the unitary layer depicted in Figure \ref{fig:IQP}. 

The resulting density matrices were obtained from the circuit and used to calculate the elements of the kernel matrix using the Hilbert-Schmidt inner product ($K(\boldsymbol{x}_i, \boldsymbol{x}_j) = \mathrm{Tr}(\rho(\boldsymbol{x}_i) \rho(\boldsymbol{x}_j))$). This quantum kernel matrix was then used to train the classical SVM model from which we computed the margin for each sample.

This experiment was conducted using $5$-fold cross-validation (CV) \cite{BERRAR2025638} for reliability, with $250$ samples for the HTRU2 dataset, $500$ samples for the Gaussian dataset and $200$ samples used for the Wine dataset. The entire Heart Disease dataset consisting of $303$ samples was also used. Each dataset was preprocessed and scaled to the range $[0, \pi]$ and transformed using principal component analysis (PCA) \cite{wold1987principal} such that the number of features in the dataset matched the number of qubits in the quantum circuit; a necessary condition for IQP encoding.

\begin{figure*}[t]
    \centering
\includegraphics[width=1\linewidth, height = 6cm]{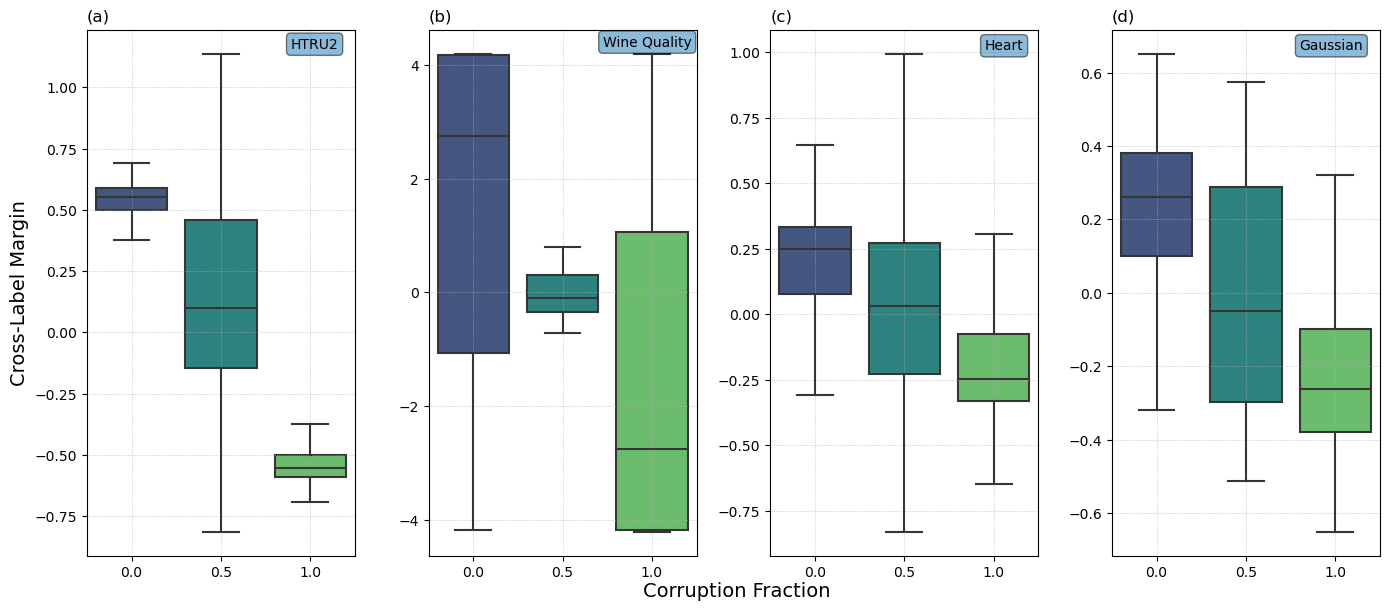}
    \caption{Box-plots depicting the cross-label margin distribution calculated using datasets with increasing fractions of corrupted training labels for various datasets. Each dataset is labelled at the top-right corner of each plot. These plots were generated without the outliers to maintain the emphasis on the median margin.}
    \label{fig:boxplots}
\end{figure*}

Figure \ref{fig:boxplots} depicts boxplots for the various datasets showing the margin distribution for increasing fractions of label corruption. These plots depict the margin distribution when no labels were corrupted (corruption fraction of $0$), when half of the training dataset had flipped labels and when all training labels were flipped (corruption levels $0.5$ and $1$, respectively).  

The plots become increasingly left-skewed as the fraction of corrupted labels increases. This can be observed since the median position of the margin distribution moves closer to more negative values as the level of corruption increases. Here, negative values indicate misclassifications of the QSVM model, which suggests poor QSVM generalisation performance.

\begin{figure*}[t]
    \centering
\includegraphics[width=1\linewidth, height = 6cm]{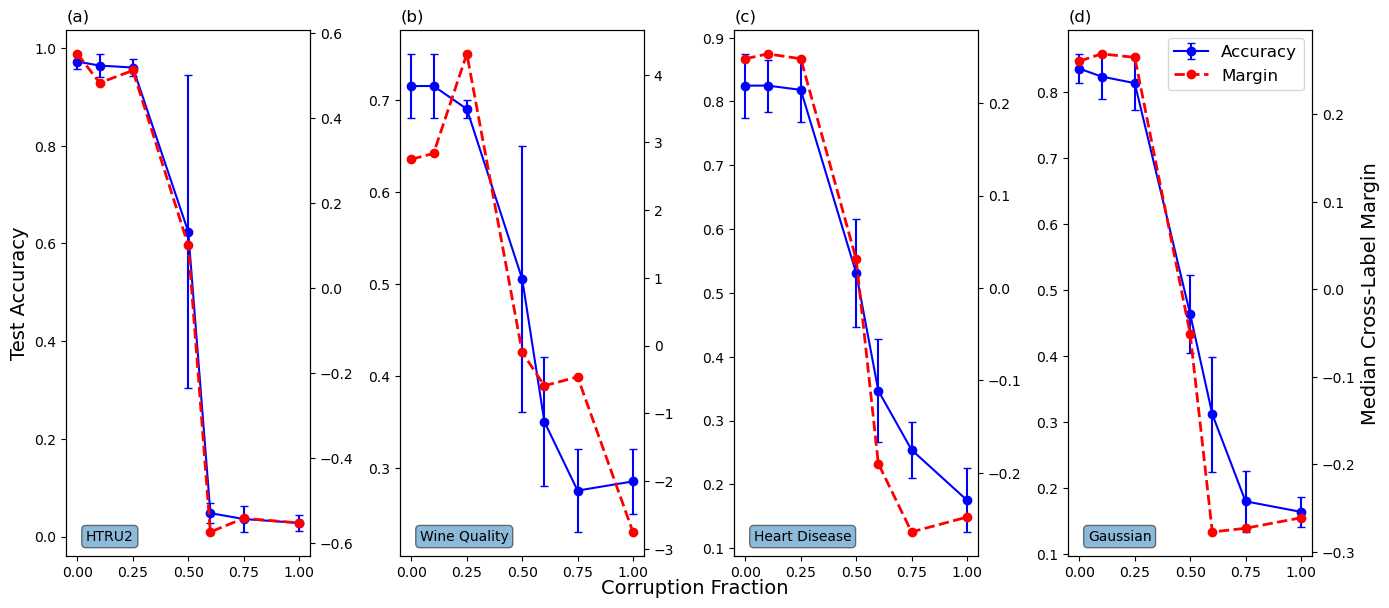}
    \caption{Plots overlaying the decreasing test accuracy and median cross-label margin graphs for increasing fractions of corrupted training labels. The test accuracy (blue) is depicted with the error obtained from 5-fold CV, and the median margins (red) are taken from the box-plots in Figure \ref{fig:boxplots} above. Each dataset is labelled at the bottom-left corner of each plot.}
    \label{fig:accMargCorr}
\end{figure*}

We further analyse this relationship in Figure \ref{fig:accMargCorr}, where we now plot the test accuracy with the median cross-label margin against increasing fractions of corrupted training labels. Here, the test accuracy was obtained from the average accuracy achieved using $5$-fold CV, with the standard deviation used as the error. This plot was constructed for the four above-mentioned datasets. It can be seen from the figure that as corruption levels increase, the shapes of the accuracy and median margin curves align almost completely (within the error bars). This is especially noted for the HTRU2 dataset.

\begin{figure*}[t]
    \centering
\includegraphics[width=1\linewidth, height = 6cm]{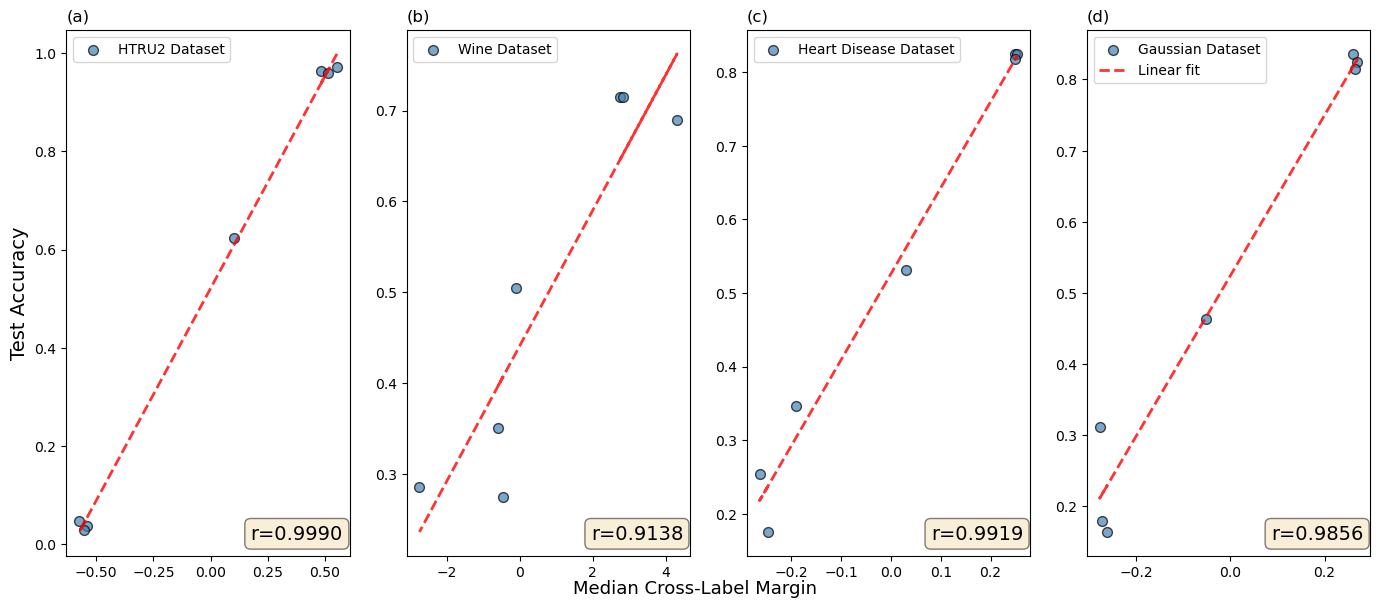}
    \caption{Plots depicting the linear correlation between test accuracy and the median cross-label margin. The Pearson correlation coefficient obtained using linear regression is depicted at the bottom-right corner of the plot for each dataset. The linear fit for each dataset is depicted by the dashed red line.}
    \label{fig:AccMargin}
\end{figure*}

Following this observation, the linear relationship between the test accuracy and the median margin, depicted in Figure \ref{fig:AccMargin}, was explored. These plots suggest an empirical correlation between the median cross-label margin and test accuracy for all four datasets within this experimental setting. Additionally, linear regression was performed for each dataset, and the Pearson correlation coefficient, $r$, was computed. This coefficient measures the linear correlation between two variables and is depicted for each dataset. The Pearson correlation coefficient can have values in the range $[-1, 1]$, with $1$ ($-1$) indicating perfectly correlated variables with an increasing (decreasing) trend.
A coefficient of $0$ suggests no relationship between the involved variables. All datasets have Pearson coefficients greater than $0.9$, with the HTRU2 dataset having the highest coefficient at $r=0.9990$, suggesting an almost perfect correlation between the variables, despite clustering at extreme corruption levels.

Our results suggest that margins are a potential indicator of generalisation performance for QSVMs. Since generalisation and robustness have been shown to be linked, this suggests that the study of margin-based robustness measures for QSVMs is worthwhile. 

However, it must be noted that this work does not include the effects of shot noise. Kernel elements are computed directly using the Hilbert-Schmidt inner product. Additionally, when using cross-validation, the test set sizes were consistently smaller than the training set sizes. It remains an open question whether the observed relationship would still hold in the face of shot noise or regimes where the unseen dataset is significantly larger.

\section{Comparing Global and Local Depolarising Noise Models}\label{sec: LocalVGlobal}

In this section, we compare the effect of increasing global and local depolarising noise on QSVM test accuracy across various datasets. This experiment is performed to motivate the use of local depolarising noise as the primary noise model considered in the numerical application of the stability bounds considered in Section \ref{sec: Results} and validated in Section \ref{sec: NumericalExp}.

In order to fairly compare the effect of global and local depolarising noise on a system of $N$ qubits, we first equate their survival probabilities.
The survival probability is the probability of an $N$-qubit state to remain unchanged after the application of a depolarising noise channel.

The survival probabilities can be easily obtained from the definitions of the depolarising noise models in Section \ref{sec: DepolNoiseModels}. From equation (\ref{globalDepol}), it is clear that for $N$ qubits, the survival probability for global depolarising noise, $p_{GS}$, is given by 
\begin{equation}\label{probSurvGlob}
    p_{GS} = (1-p_G)
\end{equation}

\noindent From equations (\ref{localDep}) and (\ref{KrausSum}), it can be shown that for $N$ qubits, the survival probability for local depolarising noise, $p_{LS}$, is expressed as

\begin{equation}\label{probSurvLoc}
    p_{LS}=(1-p_L)^N
\end{equation}

\noindent where $p_G$ ($p_L$) is the probability associated with global (local) depolarising channel over one encoding layer (qubit) and $N$ is the number of qubits.

\begin{figure*}[t]
    \centering
    \includegraphics[width=1\linewidth, height = 6cm]{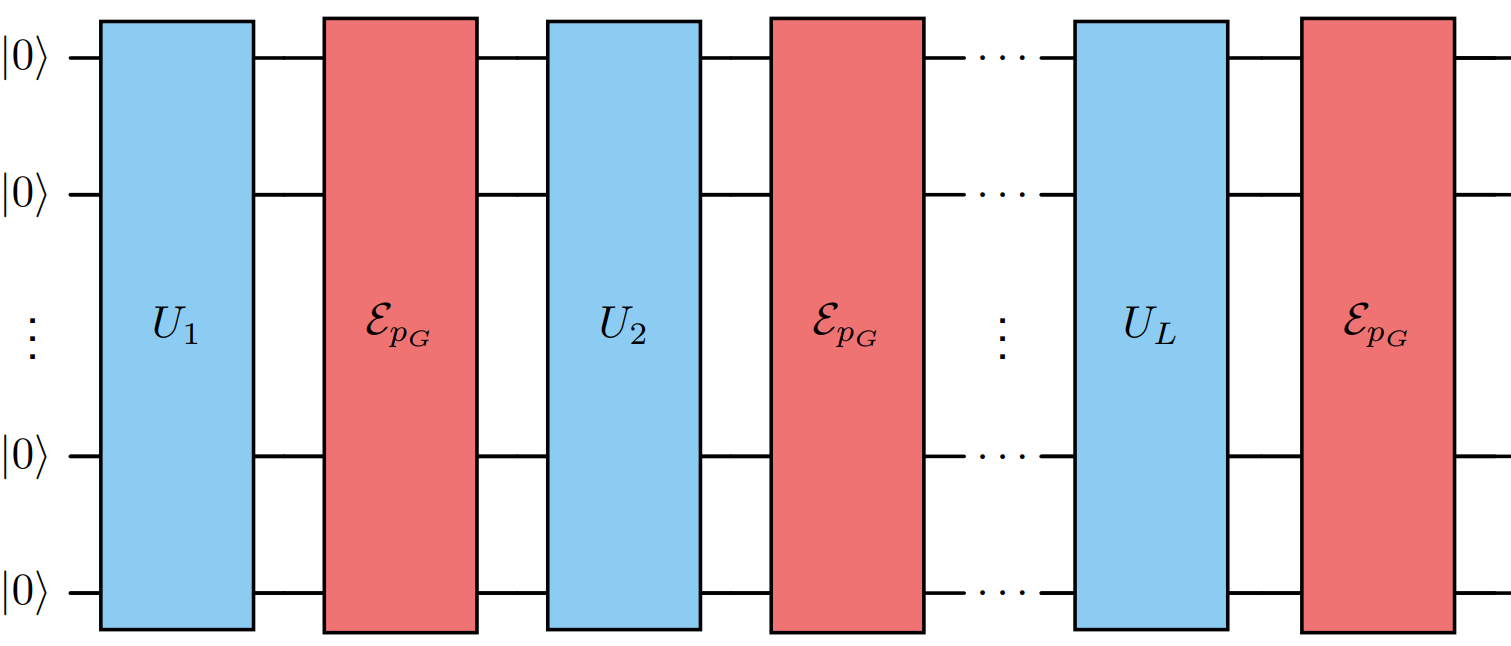}
    \caption{Quantum circuit used to implement global depolarising noise channels for $N$ qubits with $L$ unitary encoding layers (blue) followed by a layer of global depolarising noise channels (red).}
    \label{fig:LlayersCircuit}
\end{figure*}

\begin{figure*}[t]
    \centering
    \includegraphics[width=0.65\linewidth, height = 5cm]{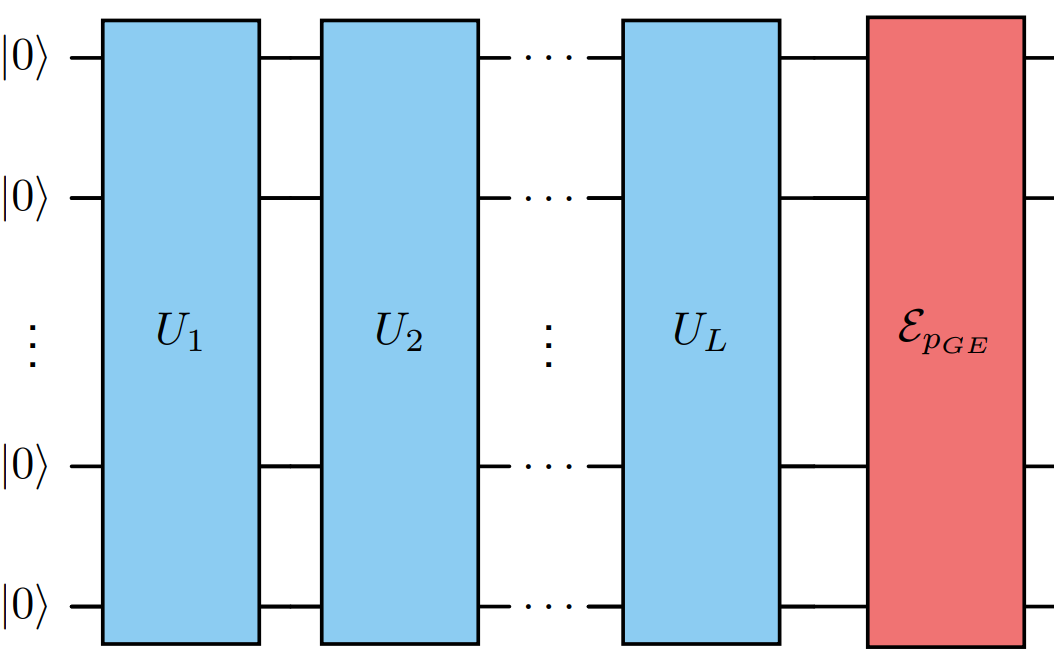}
    \caption{Equivalent quantum circuit used to implement global depolarising noise channels for $N$ qubits with $L$ unitary encoding layers (blue) followed by one layer of global depolarising noise channels (red) at the end of the circuit with a scaled global noise value. The scaling used corresponds to $p_{GE} = 1 - (1-p_G)^L$, where $p_G$ is the global noise value used per noise layer in Figure \ref{fig:LlayersCircuit}.}
    \label{fig:Equiv1layer}
\end{figure*}

Adapting the above equations for $L$ encoding layers requires the result of \citet{du2021learnability}. This result demonstrates that a circuit with $L$ layers of global depolarising channels applied after each unitary layer, as depicted in Figure \ref{fig:LlayersCircuit}, is equivalent to the circuit portrayed in Figure \ref{fig:Equiv1layer} in which one noise channel is applied at the end of the circuit with the scaling $p_{GE} = 1-(1-p_G)^L$.

From the above result, for $L$ unitary layers, equations (\ref{probSurvGlob}) and (\ref{probSurvLoc}) become $p_{GS} = (1-p_{GE})$ and $p_{LS}=(1-p_L)^{NL}$, respectively.

Henceforth, the terms local (global) test accuracy will be used to describe the accuracy obtained from the QSVM model trained using quantum kernel values obtained from a quantum circuit with local (global) depolarising noise channels. Additionally, since only depolarising noise is discussed here; local and global noise will be used instead, for brevity.

\begin{figure*}[t]
    \centering
    \includegraphics[width=1\linewidth, height = 6cm]{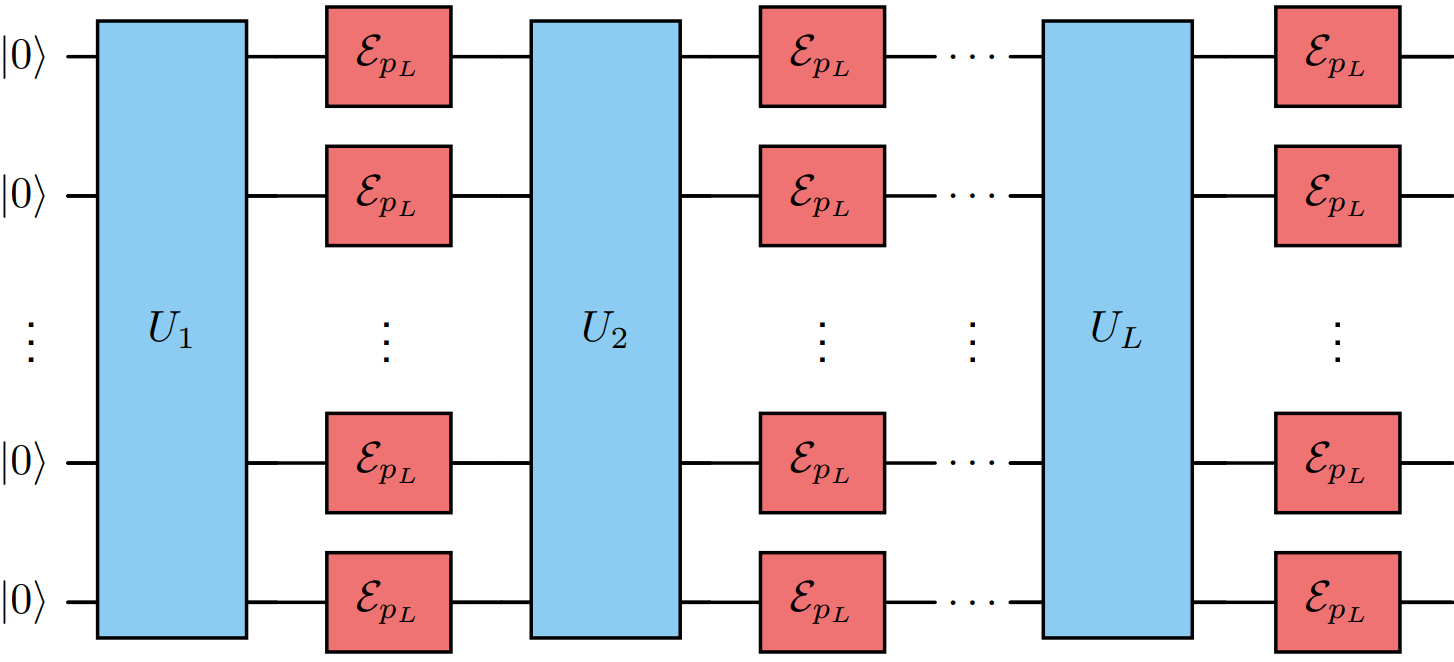}
    \caption{Quantum circuit used to implement local depolarising noise channels for $N$ qubits with $L$ unitary encoding layers (blue) followed by a layer of local depolarising noise channels (red).}
    \label{fig:LlayersLocalCircuit}
\end{figure*}

The methodology to obtain the local test accuracy follows similarly as described in Section \ref{sec: MarginGenLink}, with the quantum circuit described by Figure \ref{fig:LlayersLocalCircuit}. This circuit depicts local depolarising noise channels applied to each qubit after each unitary layer.

Following the above discussion, the survival probabilities of the noise models were equated. Thereafter, the resulting density matrix from the ideal quantum circuit was obtained. The density matrix affected by global depolarising noise was computed using equation (\ref{globalDepol}). This density matrix is then used to calculate the quantum kernel element, which will, in a similar manner, yield the global test accuracy.

\begin{figure*}[t]
    \centering    \includegraphics[width=\linewidth, height = 6.5cm]{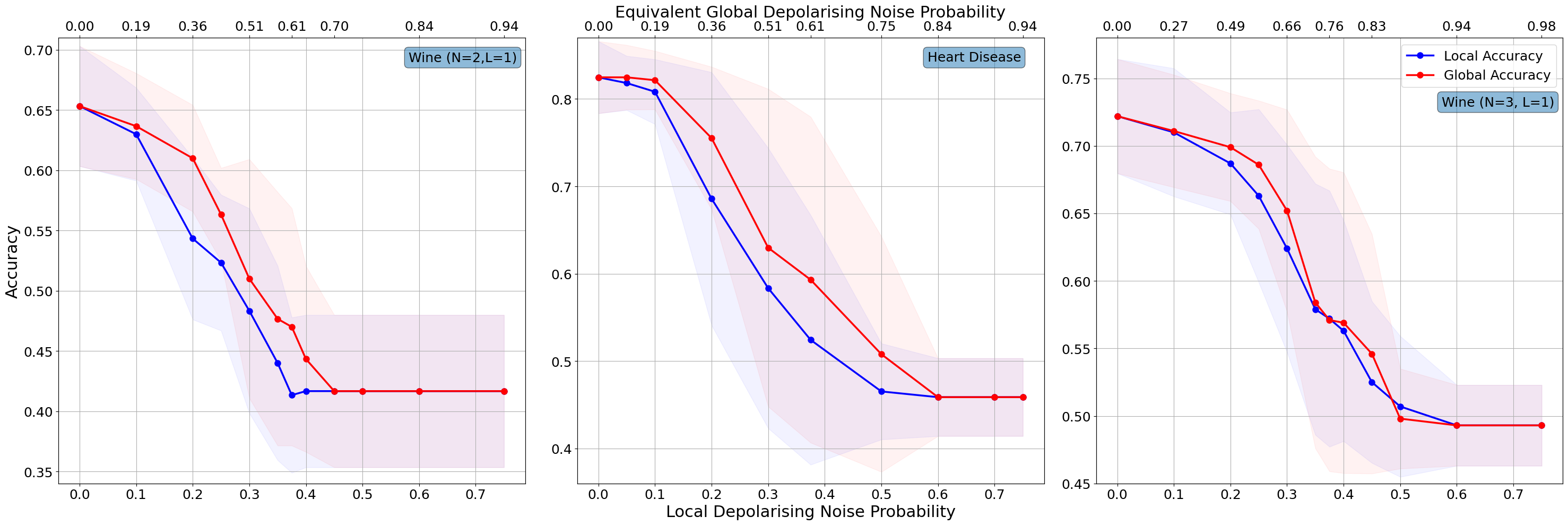}
\caption{The average test accuracies obtained after implementing 5-fold cross-validation (CV) when using the global and local depolarising noise models are depicted by the red and blue lines, respectively, for varying local noise levels. The dots are joined to guide the eye. The equivalent global noise probability is depicted for selected local noise probabilities using a second abscissa placed above the three plots for readability. The light blue and red shaded regions depict the uncertainty represented by the standard deviation. The purple shaded region represents the overlap of the uncertainty regions. The values, $N$ and $L$, for the Wine dataset indicate the differing number of qubits and unitary layers used in the quantum circuit, respectively. The Heart Disease dataset used two qubits with one unitary layer.}
    \label{fig:AccDiff}
\end{figure*}

Figure \ref{fig:AccDiff} portrays the test accuracies obtained when using the global and local noise models, respectively. A second abscissa is displayed, with one abscissa representing increasing levels of local depolarising noise, and the other showing the equivalent global depolarising noise level as a result of their matched survival probabilities. This simulation was performed for two datasets; the Wine Quality and Heart Disease datasets. The Wine Quality dataset was analysed using $1000$ samples, with two and three qubits, respectively. The Heart Disease dataset was analysed using two qubits and $303$ samples. The average accuracies obtained after implementing $5$-fold CV were used to construct the plot. The light blue and red shaded regions represent the uncertainty of each plot. The standard deviations were used to obtain the uncertainty values for the test accuracy of each noise probability. The purple shaded region represents the overlap of the uncertainty regions.

Figure \ref{fig:AccDiff} suggests that, for the datasets considered, local depolarising noise may lead to a faster degradation in QSVM test accuracy than the matched global depolarising noise model. In particular, the average test accuracies obtained under global depolarising noise appear higher than those obtained under local depolarising noise at intermediate noise levels. However, the sizeable uncertainty regions suggest that these trends should be interpreted cautiously. Nevertheless, despite these inconclusive results, the local model incorporates noise at the level of individual qubits and may capture effects not present in the global model under the chosen matching convention. For this reason, local depolarising noise is used as the main physical application for the stability-bound analysis derived in this work.

\section{Cross-Kernel Stability Bound Framework}\label{sec: Results}

In this section, we introduce the cross-kernel margin and derive stability bounds for the corresponding inverse-squared margin quantity under general kernel perturbations. This stability analysis is performed for the ridge-stabilised QSVM dual formulation discussed in Section \ref{sec:UB}. The local depolarising noise model is then applied to this framework as an application of the bounds by first deriving an expression for the quantum kernel under local depolarising noise and then using this to describe kernel perturbations in the stability bound. 

The principal quantity defined and studied in this section is the \textit{cross-kernel margin}. Unlike a conventional SVM margin, we define the cross-kernel margin as the margin obtained by evaluating the dual solution, $\boldsymbol{\tilde{\alpha}}$, learned using a noisy kernel matrix, with respect to an ideal kernel matrix. It must be noted that this quantity does not correspond to the margin of an SVM trained on the ideal kernel, nor the margin of an SVM trained using the noisy kernel. Accordingly, the cross-kernel margin considered here should be interpreted as a robustness measure, quantifying how the classifier learned from a perturbed kernel behaves when evaluated in the ideal kernel geometry.

Since the cross-kernel margin is defined as a robustness measure, bounding its variation under kernel perturbations naturally lead to stability bounds for this quantity.

\subsection{Cross-Kernel Stability Bounds}\label{sec:UB}

In this section, we derive a stability bound for the \textit{cross-kernel margin}, denoted by $\gamma_\times$, under a general kernel perturbation. Since the margin is inversely related to the RKHS norm, as mentioned in Section \ref{sec: SVMTheory}, the analysis is most naturally stated in terms of the inverse squared-margin quantities. 

First, we let 

\begin{equation}
Q = YKY \text{ and } \tilde{Q} = Y\tilde{K}Y
\end{equation}

\noindent where $Y = \text{diag}(y_1, \cdots, y_m)$ defines the diagonal matrix with training labels $\{ y_i\}_{i=1}^m$ for $m$ training samples. $K$ is the kernel matrix obtained in the ideal RKHS and $\tilde{K}$ is the perturbed kernel. 

Since the SVM dual objective is a convex optimisation problem, we employ standard results from convex optimisation theory \cite{boyd2004convex}. In particular, we consider the ridge-stabilised dual optimisation problem by replacing the quantity $Q$ with $Q_{\tau} = Q + \tau I$, where $\tau>0$ is a small regularisation parameter. This regularisation results in strong convexity of the dual objective function and therefore ensures the existence of a unique global minimiser required for the subsequent stability analysis. 

Additionally, $\boldsymbol{\alpha}$ and $\boldsymbol{\tilde{\alpha}}$ denote the solutions of the Tikhonov-stabilised dual objectives of the ideal and perturbed kernels, respectively. The application of the Tikhonov-stabilised framework is discussed further in Appendix \ref{appendixAlphaBound}.

We define  

\begin{equation}\label{qRelations}
q_0 = \boldsymbol{\alpha}^T Q \boldsymbol{\alpha} \text{ and } q_\times = \boldsymbol{\tilde{\alpha}}^T Q \boldsymbol{\tilde{\alpha}}
\end{equation}

\noindent where  $q_0$ is the inverse squared margin associated with the ideal classifier in the ideal RKHS and $q_\times$ is the corresponding cross-kernel quantity obtained by evaluating the noise-affected dual solution, $\boldsymbol{\tilde{\alpha}}$ with respect to the ideal kernel, $K$. 

The cross-kernel and ideal margins and the cross-kernel and ideal norm quantities have the following inverse relation;

\begin{equation}\label{marginDefs}
    \gamma_0^2 = \frac{1}{q_0}
    \text{ and } \gamma_\times^2 = \frac{1}{q_\times} 
\end{equation}

We begin by bounding the deviation of the cross-kernel and the ideal quantities. From the above-mentioned definitions of $q_0$ and $q_\times$, we have

\begin{align}\label{qdiff}
q_\times-q_0
&=
\boldsymbol{\tilde{\alpha}}^TQ\boldsymbol{\tilde{\alpha}}
-
\boldsymbol{\alpha}^TQ\boldsymbol{\alpha} \\
&=
\boldsymbol{\tilde{\alpha}}^TQ(\boldsymbol{\tilde{\alpha}}-\boldsymbol{\alpha})
+
(\boldsymbol{\tilde{\alpha}}-\boldsymbol{\alpha})^TQ\boldsymbol{\alpha} \\
&=
(\boldsymbol{\tilde{\alpha}}-\boldsymbol{\alpha})^TQ(\boldsymbol{\tilde{\alpha}}+\boldsymbol{\alpha}),
\end{align}

\noindent The above result is obtained by adding and subtracting the term $\boldsymbol{\tilde{\alpha}}^TQ\boldsymbol{\alpha}$ and using the symmetry of $Q$. Applying the Cauchy-Schwarz inequality \cite{watrouscs} gives

\begin{equation}\label{toBoundCS}
|q_\times - q_0| \leq \|\tilde{\boldsymbol{\alpha}}-\boldsymbol{\alpha}\|_2 \| Q (\boldsymbol{\tilde{\alpha}} + \boldsymbol{\alpha})\|_2 
\end{equation}

\noindent where $\| \cdot \|_2$ denotes the Euclidean norm for vectors and the spectral norm for matrices.

Using the result proven in Appendix (\ref{appendixAlphaBound}), 
we can bound inequality (\ref{toBoundCS}) as, 

\begin{equation}\label{perturbationBounds}
|q_\times - q_0| \leq B_q(\boldsymbol{\tilde{\alpha}},\boldsymbol{\alpha})
\end{equation}
\noindent where 
$$
B_q = \frac{\| Q (\boldsymbol{\tilde{\alpha}} + \boldsymbol{\alpha})\|_2}{\tau} \min \{ \| P \Delta K P\boldsymbol{{z}}  \|_2, \| P \Delta K P\boldsymbol{\tilde{z}} \|_2 \} 
$$

\noindent Here, $\Delta K = \tilde{K} - K$, and $\tau$ defines the regularisation parameter associated with the applied Tikhonov regularisation. Additionally, the matrix,
$$
P = I - \frac{1}{m} \boldsymbol{1}\boldsymbol{1}^T
$$ 
is the projection onto the subspace orthogonal to the all-ones vector. This projection is used to enforce the feasibility constraint that $y^T \boldsymbol{\alpha} = \boldsymbol{0}$, where $y^T = (y_1, \cdots, y_m)^T$. Additionally
$
\boldsymbol{z} = Y \boldsymbol{\alpha}$ and  $\boldsymbol{\tilde{z}} = Y \boldsymbol{\tilde{\alpha}}  
$. Further details regarding this derivation can be found in Appendix \ref{appendixAlphaBound}.

Inequality (\ref{perturbationBounds}) implies 

\begin{equation}
 q_0- B_q(\boldsymbol{\tilde{\alpha}},\boldsymbol{\alpha})
 \leq q_\times \leq q_0 + B_q(\boldsymbol{\tilde{\alpha}},\boldsymbol{\alpha})
\end{equation}

We can rewrite this in terms of the margins defined in equation (\ref{marginDef}) for the case $ q_0 > B_q(\boldsymbol{\tilde{\alpha}},\boldsymbol{\alpha})$ as,

\begin{equation}
 \frac{1}{q_0 + B_q(\boldsymbol{\tilde{\alpha}},\boldsymbol{\alpha})}
 \leq \gamma^2_\times \leq \frac{1}{q_0- B_q(\boldsymbol{\tilde{\alpha}},\boldsymbol{\alpha})} 
\end{equation}

If $B_q(\boldsymbol{\tilde{\alpha}},\boldsymbol{\alpha}) > q_0$, the lower bound on $q_\times$ becomes negative and therefore cannot be inverted to produce an upper bound on $\gamma_\times^2$. In this case, only the lower margin bound remains informative.

Since these bounds are derived using the Tikhonov-stabilised SVM dual formulation, they apply to ridge-stabilised dual objectives that remain strongly convex after the perturbation. Furthermore, since the bound given by inequality (\ref{perturbationBounds}) \textit{a posteriori} since it is evaluated using the computed dual solutions $\boldsymbol{\alpha}$ and $\boldsymbol{\tilde{\alpha}}$, these bounds are numerically validated in Section \ref{ValidBounds}. An alternative, but conservative bound to inequality (\ref{perturbationBounds}) is derived in Appendix \ref{ConservativeBounds}.

\subsection{Bound Application: Local Depolarising Noise Model}
\subsubsection{Local Depolarising Kernel Expression}

In this section, we study how kernel elements are affected by a single application of a local depolarising noise channel for an $N$-qubit system. We derive an exact expression for the kernel affected by a single layer of local depolarising noise acting independently on each of $N$ qubits.

Using equation (\ref{localDep}), the application of a single-qubit local depolarising channel on a density matrix, $\rho$ is given by,

\begin{equation}
    \mathcal{E}_L(\rho) = (1-p_L)\rho + \frac{p_L}{3}(\sigma_{x}\rho \sigma_{x} + \sigma_{y} \rho \sigma_{y} + \sigma_{z} \rho \sigma_{z})
\end{equation}
where $(\sigma_x, \sigma_y, \sigma_z)$ are the Pauli operators.

\noindent Since, any single-qubit density matrix, $\rho$, can be expressed in the Pauli basis as,

\begin{equation}
    \rho = \frac{1}{2} \sum_{\mu=0}^3 r_{\mu}{\sigma_{\mu}}
\end{equation}

\noindent where $\sigma_0 = I$, $\sigma_1 = \sigma_x$, $\sigma_2 = \sigma_y$ and $\sigma_3 = \sigma_z$, and the coefficients $r_{\mu} = \mathrm{Tr}(\rho \, \sigma_{\mu})$, the action of the depolarising channel on the Pauli basis elements is therefore 

\begin{equation}\label{channel3PauliMain}
\mathcal{E}_L(\sigma_{\mu}) = (1-p_L)\sigma_{\mu} + \frac{p_L}{3} \sum_{i \in \{ x,y,z\}} \sigma_{i}\sigma_{\mu}\sigma_{i}
\end{equation}

\noindent To evaluate this expression, we analyse the product of three general Pauli matrices of the form   
$\sigma_{i}\sigma_{j}\sigma_{i}$. To do this, we use the results proven in Appendix \ref{AppendixB} for the product of three Pauli matrices,

\begin{equation}\label{etaRelationship}
    \sigma_i \sigma_j \sigma_i = \eta_{i,j} \sigma_j
\end{equation}

\noindent where $\eta_{i,j}$ is defined as,

\[
\eta_{i,j} = 
\begin{cases}
+1, \text{if } j \in \{0, i\}\\
-1, \text{else} 
\end{cases}
\]

\noindent Using this, we can rewrite equation (\ref{channel3PauliMain}) as,

\[
\mathcal{E}_L(\sigma_{\mu}) =
\begin{cases}
    \sigma_{\mu}, \hspace{1.8cm} \mu = 0\\
    (1-\frac{4}{3}p_L)\sigma_{\mu}, \hspace{0.23cm} \mu \neq 0
\end{cases}
\]

We can therefore express this as
\begin{equation}
\mathcal{E}_L(\sigma_{\mu}) = s_{\mu}\sigma_{\mu}
\end{equation}

where the single-qubit factors are
\[
s_{\mu} = 
\begin{cases}
1, &\mu = 0 \\
1-\frac{4}{3}p_L,& \mu \neq 0 
\end{cases}
\]

\noindent For $N$-qubits, the channel acts independently on each qubit, giving

\begin{equation}
\mathcal{E}_L^{\otimes{N}}(\sigma_{\boldsymbol{\mu}}) = \left(\prod_{k=1}^N s_{\mu_k}\right)\sigma_{\boldsymbol{\mu}}
\end{equation}

\noindent where $\boldsymbol{\mu} = (\mu_1, \cdots ,\mu_N)$ and $\sigma_{\boldsymbol{\mu}} = (\sigma_{\mu_1} \otimes \cdots \otimes  \sigma_{\mu_N})$ 

Applying the depolarising channel to the $N$-qubit density matrix, $\rho_i$ gives

\begin{equation}\label{noisyrho}
 \tilde{\rho}_i = \frac{1}{2^N}\sum_{\boldsymbol{\mu}}{r}^{(i)}_{\boldsymbol{\mu}} \left(\prod_{k=1}^N s_{\mu_k}\right)\sigma_{\boldsymbol{\mu}}
\end{equation}

\noindent where $\boldsymbol{r}_{\boldsymbol{\mu}}^{(i)} = \text{Tr}(\rho_i \sigma_{\boldsymbol{\mu}})$, $\rho_i$ is the $N$-qubit density matrix and $ \sigma_{\boldsymbol{\mu}}= \sigma_{\mu_1} \otimes \cdots \otimes  \sigma_{\mu_N}$ as defined above. We can therefore determine the corresponding $ij$-th noisy kernel element via the Hilbert-Schmidt inner product for the density matrix, $\tilde{\rho}_i$ and the similarly-defined matrix, $\tilde{\rho}_j$

\begin{equation}
      \tilde{K}_{ij} = \frac{1}{2^{2N}} \sum_{\boldsymbol{\mu},\boldsymbol{\nu}}{r}^{(i)}_{\boldsymbol{\mu}} {r}^{(j)}_{\boldsymbol{\nu}}\left(\prod_{k=1}^N s_{\mu_k} s_{\nu_k}\right)\mathrm{Tr}(\sigma_{\boldsymbol{\mu}}\sigma_{\boldsymbol{\nu}})
\end{equation}

\noindent Using trace properties, we obtain 

\begin{equation}\label{ExactMain}
      \tilde{K}_{ij} = \frac{1}{2^{N}} \sum_{\boldsymbol{\mu}}{r}^{(i)}_{\boldsymbol{\mu}} {r}^{(j)}_{\boldsymbol{\mu}}\left(\prod_{k=1}^N s^2_{\mu_k}\right)
\end{equation}

This expression describes the noisy kernel for a single application of the local depolarising channel to the $N$-qubit system. The full detailed derivation of the above result can be found in Appendix \ref{AppendixB}. 
 
While it is possible to obtain a formal compact closed-form expression for the $ij$-th noisy kernel element given by equation (\ref{ExactMain}), the expression involves a sum over a combinatorially large number of terms, each weighted by different factors, making it analytically intractable for large $N$. Therefore, an auxiliary upper bound for kernel elements affected by local depolarising noise is presented in Appendix \ref{AppendixKernelFull}.

\subsubsection{Local Depolarising Cross-Kernel Bounds}\label{sec: UBLocal}
As an application of the general stability derived bounds in Section \ref{sec:UB}, we present the analytical form for the case in which the quantum kernel is perturbed by a single local depolarising noise channel. It must be noted that the stability bound itself is formulated for general kernel perturbations and therefore remains applicable for perturbed kernel matrices obtained numerically. The purpose of this section is to present the analytical expression of the stability bounds using the closed-form kernel expression corresponding to a single application of the local depolarising noise channel.

We recall the exact form of the quantum kernel affected by a single local depolarising noise channel given by equation (\ref{ExactMain}) as,

\begin{equation}
    \tilde{K}_{ij} = \frac{1}{2^{N}} \sum_{\boldsymbol{\mu}}{r}^{(i)}_{\boldsymbol{\mu}} {r}^{(j)}_{\boldsymbol{\mu}}\left(\prod_{k=1}^N s^2_{\mu_k}\right)
\end{equation}

\noindent Here, as previously defined, $\rho_i$ is the $N$-qubit density matrix, $ \sigma_{\boldsymbol{\mu}}= \sigma_{\mu_1} \otimes \cdots \otimes  \sigma_{\mu_N}$ and $\boldsymbol{r}_{\boldsymbol{\mu}}^{(i)} = \text{Tr}(\rho_i \sigma_{\boldsymbol{\mu}})$. $\boldsymbol{r}_{\boldsymbol{\mu}}^{(j)}$ is defined similarly to $\boldsymbol{r}_{\boldsymbol{\mu}}^{(i)}$.  The factors $s_{\mu_k}$ are defined as

\[
s_{\mu} = 
\begin{cases}
1, &\mu = 0 \\
1-\frac{4}{3}p_L,& \mu \neq 0 
\end{cases}
\]

The ideal kernel can be retrieved from the local depolarising noise kernel by setting $p_L = 0$. This has the form,

\begin{equation}\label{Ideal}
      {K}_{ij} = \frac{1}{2^{N}} \sum_{\boldsymbol{\mu}}{r}^{(i)}_{\boldsymbol{\mu}} {r}^{(j)}_{\boldsymbol{\mu}}
\end{equation}

This results in the following expression describing the perturbation of a quantum kernel under local depolarising noise as,

\begin{align}
\Delta K_{ij} &= \tilde{K}_{ij} - K_{ij}\\
&= \frac{1}{2^{N}} \sum_{\boldsymbol{\mu}}{r}^{(i)}_{\boldsymbol{\mu}} {r}^{(j)}_{\boldsymbol{\mu}}\left(\prod_{k=1}^N s^2_{\mu_k} - 1\right)
\end{align}

Substituting this into the stability bounds yields,

\begin{equation}\label{pertBoundLocalDep}
|q_\times - q_0| \leq B_q(\boldsymbol{\tilde{\alpha}},\boldsymbol{\alpha})
\end{equation}
\noindent where 
$$
B_q = \frac{\| Q (\boldsymbol{\tilde{\alpha}} + \boldsymbol{\alpha})\|_2}{\tau} \min \{ \| P \Delta K (p_L) P\boldsymbol{{z}}  \|_2, \| P \Delta K(p_L) P\boldsymbol{\tilde{z}} \|_2 \} 
$$
\noindent where the dependence on the local depolarising noise parameter, $p_L$, enters the bound through the noise factors $s_{\mu_k}$. 

This bound is numerically validated in Section \ref{sec: NumericalExp}. It must be noted, however, that the general stability bound given by inequality (\ref{perturbationBounds}) is independent of the perturbation model and therefore applies directly to the noisy kernel matrices used throughout numerical simulations. By contrast, the exact expression for the noisy kernel presented in this section assumes a single application of the local depolarising noise channel. Consequently, inequality (\ref{pertBoundLocalDep}) is included to illustrate the analytical form of the stability bound under local depolarising noise and is not intended to provide a closed-form description of the multiple applications of the noise channel as considered in Section \ref{sec: NumericalExp}.

\Needspace{5\baselineskip}

\section{Numerical Experiments} \label{sec: NumericalExp}

\subsection{Validating the Stability Bounds under Local Depolarising Noise}\label{ValidBounds}

In this section, we outline the pipeline followed to numerically test the stability bounds derived in Section \ref{sec:UB} across multiple datasets using the local depolarising noise model as the noise model used to induce kernel perturbations. Four different datasets were used to perform five numerical simulations, and the corresponding results are depicted in Figure \ref{fig:pertBounds}. The specific parameters used for each dataset are detailed in Table \ref{tab:DatasetSelec} in Appendix \ref{app: ExpParam}. 

To validate the stability bounds, the data from each dataset is first preprocessed and scaled as discussed in Section \ref{sec: MarginGenLink}. A similar pipeline as discussed in Section \ref{sec: MarginGenLink} is used.

The training set is passed through a quantum circuit composed of two qubits and $L$ layers of unitary gates, where each layer corresponds to a layer of IQP encoding gates. The number of unitary layers used in this quantum circuit is dataset-dependent and outlined in Table \ref{tab:DatasetSelec} in Appendix \ref{app: ExpParam}. 

All numerical simulations were conducted using \textit{Pennylane} \cite{bergholm2018pennylane}. For this experiment, the quantum circuit, depicted in Figure \ref{fig:LlayersLocalCircuit}, was used to incorporate local depolarising noise channels using Pennylane's \textit{qml.DepolarizingChannel} function. As discussed in Section \ref{sec: MarginGenLink}, noise channels are applied to each qubit after each unitary layer.

Throughout the numerical experiments, the ridge-stabilised kernel matrix considered in the theoretical analysis in Section \ref{sec:UB} was employed for the subsequent QSVM training by replacing the kernel matrix, $K$ with the $K + \tau I$, where $\tau>0$ is the corresponding regularisation parameter. This ensures consistency between the considered theoretical problem and the numerical simulations.

All experiments were conducted using the default value of the SVM regularisation parameter $C = 1$. The ridge parameter was fixed at $\tau = 0.1$, as this value was found to provide sufficient regularisation without substantially altering the geometry of the kernel matrix. For increasing levels of noise, the noisy kernel is computed and used to train the SVM, from which the noisy dual solution is obtained. The cross-kernel inverse squared-margin is then evaluated using the ideal kernel.

Using the setup described above, the bounds presented by inequality (\ref{perturbationBounds}), was tested using four different datasets. 
The observed relative deviation is calculated as the absolute difference of these two quantities, as $|q_\times - q_0|/q_0$. The corresponding bound, given by equation (\ref{perturbationBounds}) is computed for increasing levels of local depolarising noise and similarly scaled by the ideal quantity, $q_0$. 

Figure \ref{fig:pertBounds} plots the stability bounds derived in Section \ref{sec:UB}, where the observed relative deviation and theoretical bounds are depicted by blue and red lines, respectively. The maximum relative deviation between the empirical and theoretical bound values is shown by the light blue dotted line for each plot.

As depicted in Figure \ref{fig:pertBounds}, for all tested local depolarising noise values and datasets, the observed relative inverse squared-margin deviation is upper-bounded by the corresponding stability bound. A brief description of each dataset is provided in Appendix \ref{AppendixE}.

\subsection{Testing the Bound with Hardware Kernels}
Two subsets of $20$ and $45$ samples, respectively, from the Breast Cancer dataset and one subset of $35$ samples from the Gaussian dataset were used to test the stability bound on noisy kernels obtained from real quantum hardware. Kernel matrices were estimated for each subset using the ibm$\_$fez device with $10000$ shots. These subsets were also used to perform numerical simulations under increasing local depolarising noise for comparison with the hardware-derived results. The relatively small subset sizes were chosen to ensure a computationally manageable number of quantum kernel circuit evaluations while still allowing the bound to be evaluated on hardware-derived kernels.

\begin{table}[!htbp]
    \caption{Scaled observed inverse squared-margin deviations, $|q_\times - q_0|/q_0$, obtained using kernel matrices from the ibm{\_}fez device and the \textit{FakeFez} simulator are presented for various datasets. The corresponding scaled stability bound values, $B(\boldsymbol{\alpha}, \boldsymbol{\tilde{\alpha}})/q_0$ are also reported.}
    \label{tab:IBM_results}
    \begin{tabular}{>{\centering\arraybackslash}p{2.5cm} >    {\centering\arraybackslash}p{2.2cm}>
    {\centering\arraybackslash}p{2.2cm}}
    \toprule % <-- Toprule here
    \textbf{Dataset}  &  $|q_\times - q_0|/q_0$  & $B(\boldsymbol{\alpha}, \boldsymbol{\tilde{\alpha}})/q_0$ \\
      \midrule % <-- Midrule here
    Breast Cancer (IBM) & 3.84 & 107.97 \\  
      \midrule  
    Breast Cancer (FakeFez)   & 0.86 & 31.21 \\  
      \midrule  
       Gaussian (IBM)  & 2.51 & 116.67\\
        \midrule
       Gaussian (FakeFez)  & 2.51 & 119.80\\
        \midrule
      Breast Cancer (IBM)  & 2.09 & 130.05 \\ 
        \midrule
      Breast Cancer (FakeFez)  & 2.20 & 134.28\\ 
      \bottomrule % <-- Bottomrule here
    \end{tabular}
\end{table}

These hardware-derived kernels are not intended to validate the local depolarising application of the stability bounds presented in Section \ref{sec: UBLocal}. Since the effective noise acting on the real IBM device and the \textit{FakeFez} simulator is not characterised by a known local depolarising parameter, the results obtained using the hardware-derived kernels are reported separately in Table \ref{tab:IBM_results} and should be interpreted as an out-of-model consistency check of the stability framework. The corresponding numerical simulations are presented in Figure \ref{fig:IBM_Pert}.

This numerical experiment was performed using the software development kit, Qiskit \cite{qiskit2024} and the $\text{ibm\_fez}$
device \cite{ibmq}, which is one of the IBM Heron processors. Additionally, numerical simulations were also performed using the \textit{FakeFez} simulator, which is a fake $156$ qubit backend with similar calibration properties to the real $\text{ibm\_fez}$ device \cite{ibmq}. One unitary layer and two qubits were used with identical preprocessing methods as discussed in Section \ref{sec: MarginGenLink}. A unitary layer of gates corresponding to IQP encoding was similarly used to maintain consistency. In this case, however, the kernel elements were computed using a different but equivalent method known as the Inversion Test, and the kernel matrix was returned. This method was employed as the density matrices used in the previous numerical experiments are only able to be extracted in simulations, whereas kernel values on real hardware must be estimated via measurement outcomes.

\begin{figure*}[t]
\centering
\includegraphics[width=1\linewidth]{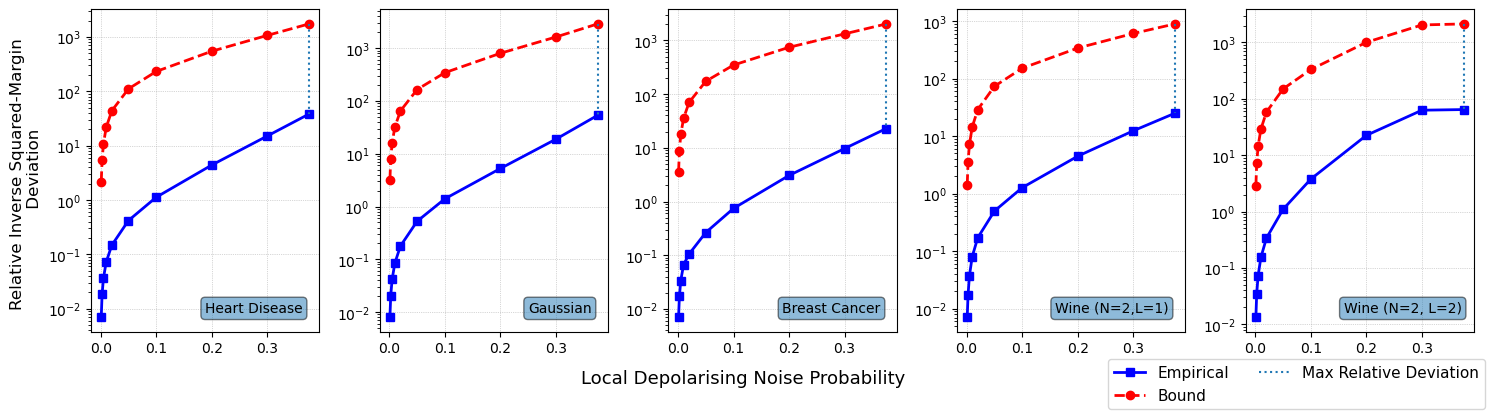}
\caption{Validation of the stability bounds derived for the cross-kernel inverse squared margin, $q_\times$ under local depolarising noise. The observed relative inverse squared-margin deviation, $|q_\times - q_0|/q_0$ is compared with the scaled theoretical bound, $B(\boldsymbol{\alpha}, \boldsymbol{\tilde{\alpha}})/q_0$ for increasing levels of local depolarising noise. All numerical simulations were performed with parameters $\tau = 0.1$ and $C=1$ for consistency. A logarithmic vertical scale is used since the derived bounds are conservative. Additionally, the ideal case, where $p=0$ was omitted from each plot due to the resulting deviation being approximately $0$, which is not informative on a logarithmic  scale.  The maximum relative deviation between the observed and theoretical values (light blue) is depicted for each plot. Each dataset is labelled at the bottom-right corner of its plot.  The values, $N$ and $L$, for the Wine dataset indicate the differing number of qubits and unitary layers used in the quantum circuit, respectively. All datasets with unspecified values used two qubits with one unitary layer.}
\label{fig:pertBounds}
\end{figure*}

The scaled deviation between the inverse of the cross-kernel and ideal margins squared, $|q_\times - q_0|/q_0$ and the corresponding theoretical bound values, $B(\boldsymbol{\alpha}, \boldsymbol{\tilde{\alpha}})/q_0$  were computed using the kernel matrices obtained from the quantum hardware and from the \textit{FakeFez} simulator. For all three subsets, the regularisation parameters, $C=1$ and $\tau=0.1$ were used in all numerical experiments for consistency with the numerical simulations. The experimental parameters used to test these bounds are presented in Table \ref{tab:IBM_expParam} in Appendix \ref{app: ExpParam}. 
\begin{figure*}[t]
   \centering
\includegraphics[width=1\linewidth]{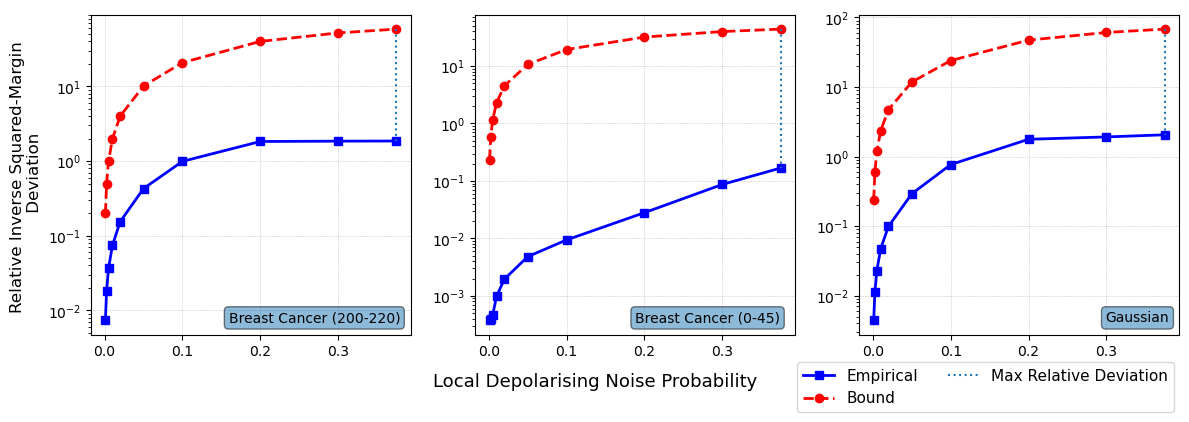}
  \caption{Local-depolarising simulations for the dataset subsets used in the hardware experiments. The scaled relative deviation of the inverse squares of the cross-kernel and ideal margins, $|q_\times - q_0|/q_0$, (blue) is compared with the scaled theoretical stability bounds, $B(\boldsymbol{\alpha}, \boldsymbol{\tilde{\alpha}})/q_0$, (red) for increasing local depolarising noise. Each plot was generated with $10000$ shots. As in Figure \ref{fig:pertBounds}, a logarithmic vertical scale is used, the ideal case, $p=0$ is omitted from the plots, and all simulations were performed using $\tau = 0.1$ and $C=1$. Two subsets of sizes $20$ and $45$ from the Breast Cancer Dataset, together with a subset of $35$ samples from the Gaussian dataset were used. Each subset is labelled at the bottom-right corner of each plot. These plots are included to allow for comparison with the hardware-derived values reported in Table \ref{tab:IBM_results}. }
    \label{fig:IBM_Pert}
\end{figure*}

Additionally, the scaled observed relative deviations with the theoretical bounds were simulated for each subset for increasing local depolarising noise similar to previous experiments depicted in Figures \ref{fig:pertBounds}. These results are presented in Figure \ref{fig:IBM_Pert} for comparison with the hardware results in Table \ref{tab:IBM_results}. The observed relative deviations and the corresponding upper bounds are depicted with blue and red lines, respectively. As in Figure \ref{fig:pertBounds}, since the theoretical bound is conservative, a vertical logarithmic scale was used and the maximum relative deviation between the observed and theoretical values is depicted by a light-blue dotted line for each plot. 

As presented in Table \ref{tab:IBM_results}, the observed relative deviation obtained using the \textit{FakeFez} simulator and the real device are both bounded above by the theoretical bounds, despite the presence of additional sources of noise not modelled in the theoretical analysis. However, it must be noted that since hardware-derived kernels are affected by finite-shot noise, they may fail to be exactly positive semidefinite. In this case, the strong-convexity assumption used to prove the theoretical bounds will still hold as long as the minimum eigenvalue of the ridge-stabilised matrix satisfies the condition $\lambda_{\min}(\tilde{Q}) > -\tau$ for the regularisation parameter, $\tau>0$ . A full eigenspectrum analysis for the hardware-derived kernels is left for future work. 

At this stage, it is important to mention that, in addition to local depolarising noise, other types of noise, such as cross-talk noise in quantum circuits, also contribute to errors and degraded predictive performance. Cross-talk noise occurs for real quantum devices when gate operations performed on a specific qubit propagate to neighbouring qubits, resulting in errors that are not explicitly captured by the depolarising noise models \cite{guan2023suppression}. While not modelled, errors due to cross-talk, dephasing noise, and shot noise may still contribute to the hardware results obtained in this work and represent an important avenue for future work.

\section{Conclusion}\label{sec: Conclusion}

In this work, we introduced the \textit{cross-kernel margin}, a margin-based robustness measure for QSVMs trained using perturbed kernel matrices. This quantity assesses the dual solution learned from a perturbed kernel with respect to the ideal kernel geometry. This metric quantifies the stability of the learned classifier relative to the ideal feature space, thereby serving as a measure of robustness for QSVMs under kernel perturbations.

We motivated the use of margin-based quantities by presenting numerical evidence indicating an empirical correlation between margin-based metrics and generalisation performance for QSVMs using corrupted labels in the noiseless regime. In particular, we show empirical correlations between the median margin value and the test accuracy obtained for four datasets using the \textit{cross-label margin}, an analogous margin-based quantity to the cross-kernel margin. A range of datasets with different properties was selected to ensure reliable results. We show empirical evidence that, for our experimental setting, the margin distribution becomes negatively skewed with an increase in training label corruption, as evidenced by the shrinking median margin value. Furthermore, we show a similar decrease in test accuracy with increasing fractions of corrupted labels for all datasets. Following this, we performed linear regression to determine the relationship between test accuracy and the median cross-label margin and observed empirical linear correlations, coupled with high Pearson correlation coefficients ($r>0.9$) for all datasets. 

These findings suggest that margin-based quantities may serve as potential empirical indicators of generalisation performance for QSVMs, motivating their use in this robustness study. However, our study does not include the effects of shot noise or other hardware-induced noise sources, such as cross-talk. Additionally, the sizes of the test sets used throughout the study were smaller than the training sets. We note that the inclusion of additional noise sources or the use of unseen datasets that are larger than the training datasets may influence the observed relationships involving the margin-based quantities considered in this study. 

We then derived a conservative, \textit{a posteriori} stability bound for the deviation between the cross-kernel and ideal inverse-squared margin quantities under kernel perturbations using the Tikhonov stabilisation framework. The cross-kernel margin is obtained by evaluating the dual solution returned by the SVM trained on a perturbed kernel matrix with respect to the ideal kernel matrix. The local depolarising noise model is then used as an application of these bounds by first deriving a closed-form expression for a quantum kernel under a single application of the local depolarising noise channel and using this in the analytical bound. The resulting stability bound was then numerically validated across multiple datasets. In this case, this quantity provides a practical measure of the robustness of a QSVM under noise. Since these bounds have been derived under the Tikhonov-stabilised SVM dual formulation, they consequently apply to perturbations for which the perturbed Tikhonov-regularised dual objective remains strongly convex.

Additionally, we motivated the use of local depolarising noise as the principal application of the bounds rather than the widely used global depolarising noise model despite its analytical intractability for large system sizes by comparing QSVM test accuracy under both models. The empirical results obtained for three datasets suggest that local depolarising noise can lead to faster degradation in test accuracy compared to the matched global depolarising model at intermediate noise levels. However, their large uncertainties must be kept in mind before making any conclusions. Despite these inconclusive results, we motivate the study of the local depolarising noise model since it is able to capture per-qubit interactions that may not be captured by the global noise model.

We conclude by presenting results from five numerical simulations to validate our stability bounds under local depolarising noise using four different datasets with varying properties. In addition to this, our results include three numerical experiments involving subsets of varying sizes from two datasets, using the $\text{ibm\_fez}$ device and the \textit{FakeFez} simulator. These results constitute consistency checks of the general stability framework using hardware-derived kernels. Future work includes applying the stability bounds to more realistic noise models, including cross-talk, dephasing and shot noise, which will allow for a more complete view of QSVM robustness on real quantum devices. 

\section*{Acknowledgements}
\noindent This research was supported by the National Research Foundation (NRF) of South Africa. The authors would like to acknowledge the National Institute for Theoretical and Computational Sciences (NITheCS). We thank the University of KwaZulu-Natal (UKZN) for the use of the HEP1 machine. We acknowledge the South African Quantum Technology Initiative (SA QuTI) and IBM Quantum services for this work. The views expressed are those of the authors, and do not reflect the official policy or position of IBM or the IBM Quantum team.

\section*{Conflict of Interest}
\noindent The authors declare no conflict of interest.

\section*{Data Availability Statement}
\noindent The Gaussian dataset was generated using scikit-learn. All other datasets used in this study were obtained from the UCI Machine Learning Repository \cite{ucimlrepo}.

\section*{Code Availability}
\noindent The code associated with this work is available at 
https://github.com/saarishag/cross${\_}$kernel${\_}$margin${\_}$nisq \cite{Govender_The_Cross-Kernel_Margin}. 
%\clearpage

\bibliographystyle{unsrtnat}
\bibliography{bibliography}

\appendix
\setcounter{section}{0}
\section*{Appendix}

\counterwithin*{equation}{section}\renewcommand\theequation{\thesection\arabic{equation}}

\counterwithin*{table}{section}\renewcommand{\thetable}{\Alph{section}\arabic{table}}
\setcounter{table}{0}

\section{Detailed Derivation of Kernel Bounds}\label{AppendixKernelFull}
In this section, we present the full derivation of the bounds for the case of kernel elements affected by a single application of a local depolarising noise channel for an $N$-qubit system.

Using the definitions in Section \ref{sec: Theory}, we employ the tensor product of the single qubit local depolarising noise channel, ($\mathcal{E}^{\bigotimes^N}_L = \mathcal{E}_L \otimes \mathcal{E}_L \otimes \cdots \otimes \mathcal{E}_L 
$) to express the resulting state after individually applying a local depolarising channel on each qubit in an $N$-qubit system as,

\begin{align}
    \tilde{\rho} = \mathcal{E}(\rho) =  
     \sum_{\boldsymbol{m} \in \{0,1,2,3\}^N } \left(\bigotimes^N_{k=1} K_{m_k} \right) \rho \left(\bigotimes^N_{k=1} K_{m_k}^{\dagger} \right)
\end{align}

\noindent where $\boldsymbol{m} = \{m_1, m_2,...m_N\}$ is the multi-index for all possible combinations of the Kraus operators defined by the expressions in (\ref{krausOp}) with $m_k\in [0,3]$ for each qubit $k$.

This can be written explicitly as, 
\begin{equation*}
    \tilde{\rho} =
     \sum_{m_1, \cdots m_N=0}^3   (K_{m_1} \otimes \cdots \otimes K_{m_N}) \, \rho \, (K_{m_1}^{\dagger} \otimes \cdots \otimes K_{m_N}^{\dagger})
\end{equation*}

\noindent For some kernel matrix of size $n \times n$, we first bound the ratio of the $ij$-th noisy kernel element with its corresponding noiseless element.

Using the Hilbert-Schmidt inner product, we get the relative expression,

\begin{equation}
    \frac{\tilde{K}_{ij}}{K_{ij}} = \frac{\mathrm{Tr}(\tilde{\rho_i} \tilde{\rho_j})}{\mathrm{Tr}(\rho_i \rho_j)}
\end{equation}

\noindent where $\tilde{\rho_i}$ represents the noisy density matrix representing the state after the application of the noisy depolarising channel. We note that this ratio is defined only when $K_{ij} \neq 0$. The case, $K_{ij} = 0$, is excluded from the analysis. 

Next, we use the expression for the noisy density matrix and apply the linearity and cyclic properties of the trace operator. Noting that the Kraus operators defined in expression \ref{krausOp} are Hermitian ($K=K^{\dagger}$), since the Pauli matrices are also Hermitian, we obtain the form,

\begin{align}
    \begin{split}\frac{\tilde{K}_{ij}}{K_{ij}}  = (1-p_L)^{2N} + 
    \frac{1}{\mathrm{Tr}(\rho_i \rho_j)}\sideset{}{'}\sum_{\substack{
    \boldsymbol{m}, \boldsymbol{n} \in \{0,1,2,3\}^N}} \\ \mathrm{Tr}\left( \rho_i \left( \bigotimes_{k=1}^N K_{m_k}K_{n_k}\right)\rho_j \left( \bigotimes_{k=1}^N K_{n_k}K_{m_k}) \right) \right)
    \end{split}
\end{align}

The first term is the result of separating the $\boldsymbol{m}= \boldsymbol{n}=0$ term from the rest of the sum and applying the definition of the Kraus operator, $K_0$ defined in equation (\ref{krausOp}) above. Hence, the primed summation, $\sideset{}{'}\sum_{\substack{\boldsymbol{m}, \boldsymbol{n} \in \{0,1,2,3\}^N}}$ sums over $\boldsymbol{m}$, $\boldsymbol{n} \in \{ 0,1,2,3\}^N$ where $\boldsymbol{m}$ and $\boldsymbol{n}$ cannot simultaneously be $\boldsymbol{0}$.

The second term captures the contributions where at least one of the summation variables $m$ and $n$ are non-zero, capturing the effect of the Kraus operators involving non-identity Pauli components. 

Now, we bound the contribution from the non-identity Pauli components, defined by $\tilde{N}$

\begin{align}
    \tilde{N} = \frac{1}{\mathrm{Tr}(\rho_i \rho_j)}\sum_{\substack{
    \boldsymbol{m}, \boldsymbol{n} \in \{0,1,2,3\}^N \\(\boldsymbol{m}, \boldsymbol{n} \neq \boldsymbol{0})}}\mathrm{Tr}\left(\rho_i Q_{mn}\rho_j Q_{nm} \right)
\end{align}

where, we define

\begin{equation}
    Q_{mn} = \bigotimes_{k=1}^N(K_{m_k}K_{n_k})
\end{equation}

Now, we note that for some $\alpha, \beta \in \{0,1,2,3\}$,

\begin{equation*}
    K_{\alpha}K_{\beta} = c_{\alpha} c_{\beta} \sigma_{\alpha} \sigma_{\beta}= c_{\alpha \beta} \sigma_{\alpha} \sigma_{\beta}
\end{equation*}

We define a piecewise function for the constants $c_{\alpha \beta} = c_{\alpha}c_{\beta}$ using the definition of the Kraus operators as,

\[
c_{\alpha\beta} = 
\begin{cases}
(1-p_L) & \text{ if } \alpha=\beta = 0 \\
\sqrt{\frac{p_L}{3}(1-p_L)} & \text{ if }  \alpha = 0, \beta=i \text{ or } \alpha = i, \beta=0 \\
\frac{p_L}{3} & \text{ if } \alpha \text{ and } \beta \in i
\end{cases} 
\]
where $i \in \{1,2,3\}$.

\noindent Since these constants ($c_{\alpha\beta})$ are real and symmetric (under the interchange of indices), we note that

\begin{equation}
    Q_{mn} = \prod^N_{k=1} c_{m_k n_k}\bigotimes^N_{k=1}(\sigma_{m_k}\sigma_{n_k}) =Q_{nm}^{\dagger} 
\end{equation}

\noindent Hence, this term can be rewritten as,

\begin{align}
\begin{split}
    \tilde{N} = \sideset{}{'}\sum_{\substack{
    \boldsymbol{m}, \boldsymbol{n} \in \{0,1,2,3\}^N}} \left(\prod^N_{k=1} (c_{m_k n_k})^2 \right)
    \frac{1}{\mathrm{Tr}(\rho_i \rho_j)}\\ \mathrm{Tr}\left(\rho_i \, \left(\bigotimes_{k=1}^N \sigma_{m_k}\sigma_{n_k}\right) \, \rho_j \, \left(\bigotimes_{k=1}^N \sigma_{n_k}\sigma_{m_k} \right)\right)
    \end{split}
\end{align}

\noindent Using the inequality $|\mathrm{Tr}(A)| \leq \|A\|_1$ \cite{kittaneh1985inequalities}, we can bound the trace term as,

\begin{equation}\label{tracePaulis}
\begin{split}
\mathrm{Tr}\left(\rho_i \, \left(\bigotimes_{k=1}^N \sigma_{m_k}\sigma_{n_k}\right) \, \rho_j \, \left(\bigotimes_{k=1}^N \sigma_{n_k}\sigma_{m_k} \right)\right) \leq \\ \Bigg\|(\rho_i \, \left(\bigotimes_{k=1}^N \sigma_{m_k}\sigma_{n_k}\right) \rho_j  \left(\bigotimes_{k=1}^N \sigma_{n_k}\sigma_{m_k} \right)\Bigg\|_1
\end{split}
\end{equation}

\noindent Using H\"{o}lders inequality for the Schatten norm \cite{kittaneh1985inequalities}, where $ \|AB\|_1 \leq \|A\|_p \|B\|_q $ such that $\frac{1}{p} + \frac{1}{q} = 1$, with $p=q=2$, we get
\begin{equation}
\begin{split}
    \Bigg\|(\rho_i \, \left(\bigotimes_{k=1}^N \sigma_{m_k}\sigma_{n_k}\right) \rho_j  \left(\bigotimes_{k=1}^N \sigma_{n_k}\sigma_{m_k} \right)\Bigg\|_1 \leq \\ 
    \|\rho_i \|_2  \, \Bigg\|\left(\bigotimes_{k=1}^N \sigma_{m_k}\sigma_{n_k}\right) \rho_j  \left(\bigotimes_{k=1}^N \sigma_{n_k}\sigma_{m_k} \right)\Bigg\|_2
    \end{split}
\end{equation}

\noindent Equivalently, this bound may also be obtained directly using the tracial matrix H\"{o}lder inequality \cite{baumgartner2011inequality}, $|\mathrm{Tr}(A^{\dagger}B)| \leq \|A\|_p \|B\|_q$, where $\|\cdot \|_p$ denotes the Schatten $p$-norm and as before, $\frac{1}{p} + \frac{1}{q} = 1$. Now, noting that since $U = \bigotimes_{k=1}^N \sigma_{m_k}\sigma_{n_k}$ and $V = \bigotimes_{k=1}^N \sigma_{n_k}\sigma_{m_k}$ are unitary matrices, we can rewrite this using the unitarily invariant property of Schatten norms \cite{watrouscs}, as

\begin{equation}
    \|\rho_i \|_2  \, \Bigg\|\left(\bigotimes_{k=1}^N \sigma_{m_k}\sigma_{n_k}\right) \rho_j  \left(\bigotimes_{k=1}^N \sigma_{n_k}\sigma_{m_k} \right)\Bigg\|_2 = \|\rho_i \|_2  \, \| \rho_j \|_2
\end{equation}

\noindent By employing the definition of the $p$-Schatten norm, $ \|A\|_p = [\mathrm{Tr}(|A|^p)]^{1/p}$, where $|A| := \sqrt{A^\dagger A}$ \cite{kittaneh1985inequalities} and noting that $\rho_i$ and $\rho_j$ are Hermitian and positive semi-definite, we can write this as 

\begin{align}
\begin{split}  
     \|\rho_i \|_2  \, \| \rho_j \|_2 = \sqrt{\mathrm{Tr}(\rho_i^2)\mathrm{Tr}(\rho_j^2)}
\end{split}
\end{align}

\noindent Using the relation involving the purity for some density matrix ($\rho$), $\frac{1}{2^N} \leq \ \mathrm{Tr}(\rho^2) \leq 1$, we can bound the above term as follows,

\begin{equation}
    \|\rho_i \|_2  \, \| \rho_j \|_2 =  \sqrt{\mathrm{Tr}(\rho_i^2)\mathrm{Tr}(\rho_j^2)} \leq 1
\end{equation}

\noindent Hence, the contribution with at least one non-zero summation variable becomes

\begin{equation}
    \tilde{N} 
    \leq \sideset{}{'}\sum_{\substack{
    \boldsymbol{m}, \boldsymbol{n} \in \{0,1,2,3\}^N}}\left(\prod^N_{k=1} (c_{m_k n_k})^2 \right)
    \frac{1}{K_{ij}}
\end{equation}

\noindent Thus, the value of the $ij$-th kernel element in the kernel matrix is bounded by 

\begin{equation}\label{kernelBounds}
    \tilde{K}_{ij} 
    \leq (1-p_L)^{2N}K_{ij} +
      \sideset{}{'}\sum_{\substack{
    \boldsymbol{m}, \boldsymbol{n} \in \{0,1,2,3\}^N}} \left(\prod^N_{k=1} (c_{m_k n_k})^2 \right)
\end{equation}

\noindent The second term of expression \ref{kernelBounds} is evaluated explicitly in Appendix \ref{NonIDPauliTerm} as $1-(1-p_L)^{2N}$. Using this result, we can simply rewrite the bound as

\begin{equation}\label{finalKernelBound_app}
    \tilde{K}_{ij}  
    \leq (1-p_L)^{2N}K_{ij} + (1-(1-p_L)^{2N})
\end{equation}

\noindent This upper-bounds the resulting kernel element after the single application of a local depolarising noise channel on $N$ qubits.

%expression for L layers removed since given without proof

\section{Non-Identity Pauli Term Evaluation (Kernel Bounds)}\label{NonIDPauliTerm}
Here, we explicitly evaluate the second term in expression (\ref{kernelBounds}) in Appendix \ref{AppendixKernelFull}.

We begin by restating the expression:
\begin{equation}
    \tilde{K}_{ij} 
    \leq (1-p_L)^{2N}K_{ij} +
      \sideset{}{'}\sum_{\substack{
    \boldsymbol{m}, \boldsymbol{n} \in \{0,1,2,3\}^N}} \left(\prod^N_{k=1} (c_{m_k n_k})^2 \right)
\end{equation}

We recall that the primed summation, $\sideset{}{'}\sum_{\substack{\boldsymbol{m}, \boldsymbol{n} \in \{0,1,2,3\}^N}}$ sums over $\boldsymbol{m}$, $\boldsymbol{n} \in \{ 0,1,2,3\}^N$ where $\boldsymbol{m}$ and $\boldsymbol{n}$ cannot simultaneously be $\boldsymbol{0}$.
\noindent To evaluate this term, we first consider the full (unrestricted) sum,
\begin{equation}
\sum_{\boldsymbol{m}, \boldsymbol{n}} \prod_{k=1}^{N} (c_{m_k n_k})^2
\end{equation}

\noindent This can be written as
\begin{equation}
\prod_{k=1}^{N} \left( \sum_{m_k=0}^{3} \sum_{n_k=0}^{3} (c_{m_k n_k})^2 \right)
\end{equation}

\noindent The primed sum is obtained by subtracting the excluded term corresponding to $\boldsymbol{m} = \boldsymbol{n} = \boldsymbol{0}$, giving
\begin{equation}
\sideset{}{'}\sum_{\boldsymbol{m}, \boldsymbol{n}} \prod_{k=1}^{N} (c_{m_k n_k})^2
=
\prod_{k=1}^{N} \left( \sum_{m_k=0}^{3} \sum_{n_k=0}^{3} (c_{m_k n_k})^2 \right)
-
(c_{00})^{2N}
\end{equation}

\noindent Using the definition of the coefficients $c_{\alpha\beta}$, we note that
\begin{equation}
\sum_{m=0}^{3} \sum_{n=0}^{3} (c_{mn})^2 = (1-p_L)^2 + 9\left(\frac{p_L}{3}\right)^2 + 2p_L({1-p_L)}= 1
\end{equation}
and
\begin{equation}
c_{00} = (1-p_L)
\end{equation}

\noindent Therefore, the second term simplifies to
\begin{equation}
\sideset{}{'}\sum_{\boldsymbol{m}, \boldsymbol{n}} \prod_{k=1}^{N} (c_{m_k n_k})^2
=
1^{N} - (1-p_L)^{2N}
=
1 - (1-p_L)^{2N}
\end{equation}

\section{Closed-Form Noisy Kernel Expression}
\label{AppendixB}

We derive an exact expression for the kernel affected by a single layer of local depolarising noise acting independently on each of $N$ qubits.

Using equation (\ref{localDep}), the application of a single-qubit local depolarising channel on a density matrix, $\rho$ is given by,

\begin{equation}
    \mathcal{E}_L(\rho) = (1-p_L)\rho + \frac{p_L}{3}(\sigma_{x}\rho \sigma_{x} + \sigma_{y} \rho \sigma_{y} + \sigma_{z} \rho \sigma_{z})
\end{equation}
where $(\sigma_x, \sigma_y, \sigma_z)$ are the Pauli operators.

\noindent Any single-qubit density matrix, $\rho$, can be expressed in the Pauli basis as,

\begin{equation}
    \rho = \frac{1}{2} \sum_{\mu=0}^3 r_{\mu}{\sigma_{\mu}}
\end{equation}

\noindent where $\sigma_0 = I$, $\sigma_1 = \sigma_x$, $\sigma_2 = \sigma_y$ and $\sigma_3 = \sigma_z$, and the coefficients $r_{\mu} = \mathrm{Tr}(\rho \, \sigma_{\mu})$.

The action of the depolarising channel on the Pauli basis elements is therefore 

\begin{equation}\label{channel3Pauli}
\mathcal{E}_L(\sigma_{\mu}) = (1-p_L)\sigma_{\mu} + \frac{p_L}{3} \sum_{i \in \{ x,y,z\}} \sigma_{i}\sigma_{\mu}\sigma_{i}
\end{equation}

\noindent To evaluate this expression, we analyse the product of three general Pauli matrices of the form   
$\sigma_{i}\sigma_{j}\sigma_{i}$

We use the following result for the product of two Pauli matrices,
\begin{equation}\label{twoPaulis}
    \sigma_i \sigma_j = \delta_{ij}\, I \, + i \, \epsilon_{ijk} \sigma_k
\end{equation}

\noindent where $\delta_{ij}$ is the Kronecker delta, which has the definition,

\[
\delta_{ij} = 
\begin{cases}
    1, \mathrm{if} \, i=j\\
    0, \mathrm{else}
\end{cases}
\]

\noindent and $\epsilon_{ijk}$ is the Levi-Civita symbol, which is defined as,

\[
\epsilon_{ijk} = 
\begin{cases}
+1, \text{ for even permutations of $(i,j,k)$}\\
-1, \text{ for odd permutations of $(i,j,k)$}\\
0, \text{   for repeated indices}
\end{cases}
\]

\noindent We use equation (\ref{twoPaulis}) to determine the product of three Pauli matrices (where $\sigma_i \neq \sigma_j$) to get

\begin{equation*}
\sigma_{i}\sigma_{j}\sigma_{i} = - i \, (\sigma_i \sigma_k)
\end{equation*}

\noindent Applying equation (\ref{twoPaulis}) again yields,

\begin{equation*}
\sigma_{i}\sigma_{j}\sigma_{i}  = - \sigma_j
\end{equation*}

\noindent However, when $\sigma_i = \sigma_j$, this easily simplifies to 
\begin{equation} 
    \sigma_i \sigma_j \sigma_i = \sigma_i (\sigma_i)^2 = \sigma_i (I) = \sigma_i 
\end{equation}

\noindent Using the above results, we can now introduce a generalised result for the product of three Pauli matrices,

\begin{equation}\label{etaRelationship}
    \sigma_i \sigma_j \sigma_i = \eta_{i,j} \sigma_j
\end{equation}

\noindent where $\eta_{i,j}$ is defined as,

\[
\eta_{i,j} = 
\begin{cases}
+1, \text{if } j \in \{0, i\}\\
-1, \text{else} 
\end{cases}
\]

From this, it follows that
\[\sum_{i \in \{ x,y,z\}} \sigma_i \sigma_{\mu} \sigma_i = 
\begin{cases}
 3\sigma_{0}, \mu = 0\\
-\sigma_{\mu}, \mu \neq 0
\end{cases}
\]

\noindent We can now rewrite equation (\ref{channel3Pauli}) using this relationship as,

\[
\mathcal{E}_L(\sigma_{\mu}) =
\begin{cases}
    \sigma_{\mu}, \hspace{1.8cm} \mu = 0\\
    (1-\frac{4}{3}p_L)\sigma_{\mu}, \hspace{0.1cm} \mu \neq 0
\end{cases}
\]

We can therefore define the single-qubit factors
\begin{equation}
\mathcal{E}_L(\sigma_{\mu}) = s_{\mu}\sigma_{\mu}
\end{equation}

where we define the single-qubit noise suppression factors
\[
s_{\mu} = 
\begin{cases}
1, &\mu = 0 \\
1-\frac{4}{3}p_L,& \mu \neq 0 
\end{cases}
\]

\noindent For $N$-qubits, the channel acts independently on each qubit, giving

\begin{equation}
\mathcal{E}_L^{\otimes{N}}(\sigma_{\boldsymbol{\mu}}) = \left(\prod_{k=1}^N s_{\mu_k}\right)\sigma_{\boldsymbol{\mu}}
\end{equation}

\noindent where $\boldsymbol{\mu} = (\mu_1, \cdots ,\mu_N)$ and $\sigma_{\boldsymbol{\mu}} = (\sigma_{\mu_1} \otimes \cdots \otimes  \sigma_{\mu_N})$ 

The $N$-qubit density matrix can be expressed as
\begin{equation}
 \rho = \frac{1}{2^N} \sum_{\boldsymbol{\mu}\in \{0,1,2,3\}^N}{r}_{\boldsymbol{\mu}} \sigma_{\boldsymbol{\mu}} 
\end{equation}
where ${r}_{\boldsymbol{\mu}} = \mathrm{Tr}(\rho\sigma_{\boldsymbol{\mu}})$.

Applying the depolarising channel to the density matrix, $\rho_i$ gives

\begin{equation}\label{noisyrho}
 \tilde{\rho}_i = \frac{1}{2^N}\sum_{\boldsymbol{\mu}}{r}^{(i)}_{\boldsymbol{\mu}} \left(\prod_{k=1}^N s_{\mu_k}\right)\sigma_{\boldsymbol{\mu}}
\end{equation}

The $ij$-th noisy kernel is determined via the Hilbert-Schmidt inner product

\begin{equation}
    \tilde{K}_{ij} = \mathrm{Tr}(\tilde{\rho}_i \tilde{\rho}_j)
\end{equation}

Substituting equation (\ref{noisyrho}) for $\tilde{\rho}_i$ and the analogous expression for $\tilde{\rho}_j$, we get

\begin{equation}
      \tilde{K}_{ij} = \frac{1}{2^{2N}} \sum_{\boldsymbol{\mu},\boldsymbol{\nu}}{r}^{(i)}_{\boldsymbol{\mu}} {r}^{(j)}_{\boldsymbol{\nu}}\left(\prod_{k=1}^N s_{\mu_k} s_{\nu_k}\right)\mathrm{Tr}(\sigma_{\boldsymbol{\mu}}\sigma_{\boldsymbol{\nu}})
\end{equation}

\noindent Using the following trace property:
\begin{equation}
\mathrm{Tr} (\sigma_{\boldsymbol{\mu}} \sigma_{\boldsymbol{\nu}}) = 2^N \delta_{\boldsymbol{\mu}\boldsymbol{\nu}}
\end{equation}

\noindent we obtain 

\begin{equation}\label{Exact}
      \tilde{K}_{ij} = \frac{1}{2^{N}} \sum_{\boldsymbol{\mu}}{r}^{(i)}_{\boldsymbol{\mu}} {r}^{(j)}_{\boldsymbol{\mu}}\left(\prod_{k=1}^N s^2_{\mu_k}\right)
\end{equation}

 In order to evaluate equation (\ref{Exact}), we would have to sum over $4^N$ Pauli strings, which becomes computationally intractable for large $N$. However, since the product involving the noise suppression factors, $\prod_{k=1}^N s^2_{\mu_k}$ decay exponentially with an increase in the number of non-identity Pauli operators, these terms contribute negligibly to the overall sum. Here, we propose a method of numerical approximation to evaluate equation (\ref{Exact}) by only incorporating a few terms involving non-identity Pauli operators and truncating the rest. This method will thus reduce the number of terms to evaluate in the sum, yielding an efficient numerical approximation for large $N$.

\section{Derivation of the Dual Solution Deviation Bound}\label{appendixAlphaBound}

In this appendix, we derive a bound for the Euclidean norm of the difference between the noisy and ideal dual solutions, $\boldsymbol{\tilde{\alpha}} \text{ and } \boldsymbol{\alpha}$, respectively. 

We recall the definitions

$$    Q=YKY
\text{ and }
    \tilde Q=Y\tilde K Y.
$$

\noindent where $Y = \text{diag}(y_1,\ldots,y_m)$ is the diagonal matrix of training labels and $K$ and $\tilde{K}$ are the ideal and noise-affected kernel matrices, respectively.

The standard SVM dual optimisation problem is given by

$$
\max_{\boldsymbol{\alpha}} \sum_{i=1}^m \alpha_i - \frac{1}{2}\sum_{i,j=1}^m \alpha_i \alpha_j y_i y_j K( \boldsymbol{x}_i, \boldsymbol{x}_j)
$$

\noindent subject to the constraints for $m$ training samples

$$ 0 \leq \alpha_i \leq C \text{ and } \sum_{i=1}^m \alpha_i y_i = 0$$

This can be equivalently expressed as a \textit{minimisation}  problem by minimising the negated dual objective:

$$
\min_{\boldsymbol{\alpha}} \left(  \frac{1}{2}\sum_{i,j=1}^m \alpha_i \alpha_j y_i y_j K( \boldsymbol{x}_i, \boldsymbol{x}_j) - \sum_{i=1}^m \alpha_i \right)
$$

In vector form, this becomes 

$$
\min_{\boldsymbol{\alpha}} \left(\frac{1}{2}\boldsymbol{\alpha}^T YKY \boldsymbol{\alpha} - \boldsymbol{1}^T \alpha \right)$$

\noindent where $\boldsymbol{1}$ denotes the vector with all components equal to $1$. 

As discussed in the main text, we apply Tikhonov regularisation (\cite{tikhonov1943stability,boyd2004convex}), resulting in two ridge-stabilised dual objectives of the form

$$
f(\alpha) = 
\frac{1}{2}\boldsymbol{\alpha}^T Q_\tau\boldsymbol{\alpha} - \boldsymbol{1}^T \alpha \text{ and } 
\tilde{f}(\alpha) = \frac{1}{2}\boldsymbol{\alpha}^T \tilde{Q}_\tau\boldsymbol{\alpha} - \boldsymbol{1}^T \alpha
$$

\noindent Here, we have $Q_{\tau} = Q + \tau I$ and $\tilde{Q}_\tau = \tilde{Q} + \tau I$ where $\tau>0$ is the Tikhonov stabilisation parameter, $I$ is the identity matrix and $Q=YKY$ as defined above.

Let the solutions to these optimisation problems be 

$$
\boldsymbol{\alpha}^* = \text{argmin}_{
\boldsymbol{\alpha} \in \mathcal{A}} f(
\boldsymbol{\alpha}) \text{ and } \boldsymbol{\tilde{\alpha}}^* = \text{argmin}_{
\boldsymbol{{\alpha}} \in \mathcal{A}} \tilde{f}(
\boldsymbol{\alpha}) 
$$ 

\noindent where $\mathcal{A} := \{ \boldsymbol{\alpha} \in \mathbb{R}^m: 0 \leq \alpha_i \leq C, y^T \boldsymbol{\alpha} = {0} \}$ defines the feasible set of solutions for the variable $\alpha$

The positive-semidefinite (PSD) property of kernel matrices suggests that the matrices $Q$ and $\tilde{Q}$ are also PSD ($Q \succeq 0$ and $\tilde{Q} \succeq 0$, where the symbol $\succeq$ corresponds to the Loewner order and is used to compare Hermitian matrices \cite{lowner1934monotone, boyd2004convex}).

Since  $Q$ and $\tilde{Q}$ are PSD matrices, adding the regularisation term, $\tau I$, ensures that both objective functions are strongly-convex on $\mathcal{A}$, with the strength of the convexity determined by $\tau$ \cite{boyd2004convex}. 

This can be seen as their associated Hessian matrices \cite{boyd2004convex} ($\nabla^2 f(\alpha) =  Q_\tau$ and $\nabla^2 \tilde{f}(\alpha) =  \tilde{Q}_\tau$) are both positive definite. A matrix, $H$ is positive definite if $\boldsymbol{v}^TH\boldsymbol{v} > 0$ $\forall \boldsymbol{v} \neq \boldsymbol{0}$ for some vector, $\boldsymbol{v}$. Further, a function, $f(x)$ is considered to be strongly-convex, if there exists some $\lambda > 0$ such that $\nabla^2f(x) \succeq \lambda I$ for all points, $x$ in some set.

For a differentiable convex function $f$ over a closed convex set $\mathcal{A}$, a point $\boldsymbol{\alpha}^*$ minimises $f$ over $\mathcal{A}$ if it satisfies the variational inequality (VI) \cite{kinderlehrer2000introduction}, of the form  

$$
\langle \nabla f(\boldsymbol{\alpha^*}), \boldsymbol{\alpha} - \boldsymbol{\alpha}^* \rangle \geq 0  \, \forall \boldsymbol{\alpha} \in \mathcal{A}
$$

where $C$ defines the SVM box-constraint parameter.

Applying this to the ridge-stabilised dual objectives, $f$ and $\tilde{f}$ results in 

$$
\langle Q_{\tau} \boldsymbol{\alpha^*} - \mathbf{1}, \boldsymbol{\alpha} - \boldsymbol{\alpha}^* \rangle \geq 0  \text{     } \forall \boldsymbol{\alpha} \in \mathcal{A}
$$

$$
\langle \tilde{Q}_{\tau} \boldsymbol{\tilde{\alpha}^*} - \mathbf{1}, \boldsymbol{\alpha} - \boldsymbol{\tilde{\alpha}}^* \rangle \geq 0  \text{       } \forall \boldsymbol{{\alpha}} \in \mathcal{A}
$$

Since $\boldsymbol{\alpha}$ represents a feasible solution from the set $\mathcal{A}$, we choose $\boldsymbol{\alpha} = \boldsymbol{\tilde{\alpha}^*}$ for the first inequality and $\boldsymbol{\alpha} = \boldsymbol{{\alpha}^*}$ for the second. 

This results in

\begin{equation}\label{VIs}
\langle Q_{\tau} \boldsymbol{\alpha^*} - \mathbf{1},\boldsymbol{d}\rangle \geq 0  \text{ and }
\langle \tilde{Q}_{\tau} \boldsymbol{\tilde{\alpha}^*} - \mathbf{1}, -\boldsymbol{d}\rangle \geq 0  
\end{equation}

for $\boldsymbol{d} := \boldsymbol{\tilde{\alpha}^*} - \boldsymbol{{\alpha}}^* $.

Adding these inequalities yields

$$
\boldsymbol{d}^TQ_{\tau}\boldsymbol{d} \leq -\boldsymbol{d}^TE \boldsymbol{\tilde{\alpha}^*}
$$

where $E = \tilde{Q}_{\tau} - Q_{\tau} = \tilde{Q} - Q$.

We can now use the property that
if $A \succeq B$ then $A-B$ is a PSD matrix, or $\boldsymbol{d}^TA\boldsymbol{d} \geq \boldsymbol{d}^TB\boldsymbol{d}$ for some vector, $\boldsymbol{d}$.

Since $Q_{\tau} = Q+ \tau I$ and $Q \succeq 0$, it follows that $Q_{\tau} \succeq \tau I$. This implies that 

$$
\tau \|\boldsymbol{d}\|^2_2\leq \boldsymbol{d}^TQ_{\tau}\boldsymbol{d} $$

Applying the Cauchy-Schwarz inequality \cite{watrouscs} gives us

\begin{equation}\label{tauIneqCS}
\tau \|\boldsymbol{d}\|^2_2\leq
|\boldsymbol{d}^TE \boldsymbol{\tilde{\alpha}^*}| \leq \|d\|_2 \| E \boldsymbol{\tilde{\alpha}^*} \|_2
\end{equation}

We recall here that 

$$E = \tilde{Q} - Q = Y (\tilde{K} - K) Y = Y \Delta K Y$$.

Here, we introduce the transformed variables $\boldsymbol{\tilde{z}} := Y \boldsymbol{\tilde{\alpha}^*}$  and $\boldsymbol{u} := Y \boldsymbol{d}$. Since both $ \boldsymbol{{\alpha}^*}$ and $ \boldsymbol{\tilde{\alpha}^*}$ are feasible dual solutions, they satisfy the constraints
$\boldsymbol{y}^T \boldsymbol{\alpha^*} = {0}$ and $y^T \boldsymbol{\tilde{\alpha}^*} = {0}$. Consequently, we have $\boldsymbol{y}^T \boldsymbol{d} = {0}$

After applying the label transformation, these constraints become

$$\boldsymbol{1}^T\boldsymbol{\tilde{z}} = \boldsymbol{1}^TY\boldsymbol{\tilde{\alpha}^*} = \boldsymbol{y}^T \boldsymbol{\tilde{\alpha}^*}={0}$$ and similarly $$\boldsymbol{1}^T\boldsymbol{{u}} = \boldsymbol{1}^TY\boldsymbol{{d}} = \boldsymbol{y}^T \boldsymbol{{d}} ={0}$$  

Therefore, after the transformation $\boldsymbol{\tilde{z}} = Y \boldsymbol{\tilde{\alpha}^*}$, the original feasibility constraint $\boldsymbol{y}^T \boldsymbol{\tilde{\alpha}^*}={0}$ becomes $\boldsymbol{1}^T\boldsymbol{\tilde{z}} =0$. Therefore, it follows that every transformed feasible vector $\boldsymbol{\tilde{z}}$ lies in the centred subspace orthogonal to the all-ones vector. 

Let 

\begin{equation}\label{projMatrix}
    P = I - \frac{1}{m} \boldsymbol{1} \boldsymbol{1}^T
\end{equation}

\noindent 
denote the orthogonal projector onto this centred subspace, where, as before, $I$ is the identity matrix, $\boldsymbol{1}$ represents the all-ones vector and $m$ is the number of training samples. Since $\boldsymbol{u}$ and $\boldsymbol{\tilde{z}}$ already lie in this subspace, we have 
 
$$P\boldsymbol{u} = \boldsymbol{u} \text{ and } P\boldsymbol{\tilde{z}} = \boldsymbol{\tilde{z}}$$

 Combining this with the definition, $\Delta K := \tilde{K} - K $, we can write,
\[
\begin{aligned}
\boldsymbol{d}^TE \boldsymbol{\tilde{\alpha}^*} &= \boldsymbol{d}^T Y \Delta K Y\boldsymbol{\tilde{\alpha}^*}\\
&=(Y\boldsymbol{d})^T \Delta K (Y\boldsymbol{\tilde{\alpha}^*})
\\
&=\boldsymbol{u}^T \Delta K \boldsymbol{\tilde{z}}\\
&=\boldsymbol{u}^TP \Delta K P\boldsymbol{\tilde{z}}
\end{aligned}
\]

Consequently, only the component of the kernel perturbation acting on the centred feasible subspace contributes to the analysis. 

By Cauchy-Schwarz, 

\begin{equation}  
| \boldsymbol{d}^TE \boldsymbol{\tilde{\alpha}^*} | \leq \| \boldsymbol{u} \|_2 \| P \Delta K P\boldsymbol{\tilde{z}} \|_2 
\end{equation}

Since $Y$ is a diagonal matrix with entries $\pm 1$, it preserves Euclidean norms, so

$$\|\boldsymbol{u}\|_2 = \|Y\boldsymbol{d}\|_2 = \|\boldsymbol{d}\|_2$$

Combining this with inequality (\ref{tauIneqCS}) results in 

\begin{equation}  
\tau \|\boldsymbol{d}\|^2_2 \leq | \boldsymbol{d}^TE \boldsymbol{\tilde{\alpha}^*} | \leq \| \boldsymbol{d} \|_2 \| P \Delta K P\boldsymbol{\tilde{z}} \|_2 
\end{equation}

For $\|\boldsymbol{d}\|_2 \neq 0$, we have

\begin{equation} \label{bound1}
\|\boldsymbol{d}\|_2  \leq \frac{1}{\tau} \| P \Delta K P\boldsymbol{\tilde{z}} \|_2 
\end{equation}

\noindent whereas, the inequality is trivially true for $\|\boldsymbol{d}\|_2 = 0$.

We can obtain a similar bound using the second (noisy) VI defined in expression (\ref{VIs}). By first rewriting the first (clean) VI using $ Q_\tau = \tilde{Q}_\tau - E$ and adding it to the second VI, we obtain the following result

$$
\boldsymbol{d}^T \tilde{Q}_{\tau}\boldsymbol{d} \leq -\boldsymbol{d}^TE \boldsymbol{{\alpha}^*}
$$

Similarly, by using the definition $\boldsymbol{u} = Y \boldsymbol{d}$ defining $\boldsymbol{z} = Y \boldsymbol{\alpha^*}$,and following the same procedure as outlined to obtain \ref{bound1}, we obtain 
\[
\begin{aligned}
\boldsymbol{d}^TE \boldsymbol{{\alpha}^*} &= \boldsymbol{d}^T Y \Delta K Y\boldsymbol{{\alpha}^*}\\
&=(Y\boldsymbol{d})^T \Delta K (Y\boldsymbol{{\alpha}^*})
\\
&=\boldsymbol{u}^T \Delta K \boldsymbol{{z}}\\
&=\boldsymbol{u}^TP \Delta K P\boldsymbol{{z}}
\end{aligned}
\]
\noindent where we used $$\boldsymbol{1}^T\boldsymbol{u} = \boldsymbol{1}^TY \boldsymbol{d} = \boldsymbol{y}^T\boldsymbol{d} = {0}$$
and  
$$\boldsymbol{1}^T\boldsymbol{z} = \boldsymbol{1}^TY \boldsymbol{\alpha^*} = \boldsymbol{y}^T\boldsymbol{\alpha^*} = {0}
$$

since both $\boldsymbol{\alpha^*}$ and $\boldsymbol{d}$ satisfy the constraints $\boldsymbol{y}^T\boldsymbol{\alpha^*} = {0}$ and $ \boldsymbol{y}^T\boldsymbol{d} = {0}$. To arrive at the last equality, we then used the consequent results

$$P\boldsymbol{u} = \boldsymbol{u} \text{ and } P\boldsymbol{{z}} = \boldsymbol{{z}}$$

Using $\tilde{Q}_{\tau} \succeq \tau I$, we obtain the result

$$
\tau \|\boldsymbol{d}\|^2_2\leq \boldsymbol{d}^T\tilde{Q}_{\tau}\boldsymbol{d} 
$$

Similarly, using these results with Cauchy-Schwarz and by applying the property of the diagonal matrix, $Y$, we get another bound of the form

\begin{equation} \label{bound2}
\|\boldsymbol{d}\|_2\leq  \frac{1}{\tau} \| P\Delta K P\boldsymbol{{z}} \|_2
\end{equation}

Combining results (\ref{bound1}) and (\ref{bound2}), and recalling that $\boldsymbol{d}=\boldsymbol{\tilde{\alpha}}^* - \boldsymbol{\alpha}^*$ gives us

\begin{equation}\label{pertBound}
\| \boldsymbol{\tilde{\alpha}} - \boldsymbol{\alpha}\|_2 \leq \frac{1}{\tau} \min \{  \| P\Delta K P\boldsymbol{{z}} \|_2,  \| P\Delta K P\boldsymbol{\tilde{z}} \|_2
 \} 
\end{equation}

\noindent where the superscript $^*$ denoting the optimiser has been dropped for notational simplicity. Hence, in the final expression, $\boldsymbol{{z}} = Y \boldsymbol{{\alpha}}$ and $\boldsymbol{\tilde{z}} = Y \boldsymbol{\tilde{\alpha}}$.

\section{Conservative Bound}\label{ConservativeBounds}

This appendix presents a looser, more conservative bound to the bound presented by equation (\ref{perturbationBounds}) in the main text.

This derivation follows similarly from the derivation presented in Appendix \ref{appendixAlphaBound}.

Recalling the definitions
$\Delta K = \tilde K-K$ and $P=I-\frac{1}{m}\mathbf{1}\mathbf{1}^T$, we define

\begin{equation}
    \varepsilon_\perp := \|P \Delta K P \|_2
\end{equation}
\noindent This quantity measures the size of the kernel perturbation projected onto the centred subspace induced by P. 

From Appendix \ref{appendixAlphaBound}, we have the bound

$$\| \boldsymbol{\tilde{\alpha}} - \boldsymbol{\alpha}\|_2 \leq \frac{1}{\tau} \min \{\| P \Delta K P\boldsymbol{{z}} \|_2 , \| P \Delta K P\boldsymbol{\tilde{z}} \|_2  \} 
$$
\noindent where $\boldsymbol{z} = Y \boldsymbol{\alpha}$ and $\tilde{\boldsymbol{z}} = Y \tilde{\boldsymbol{\alpha}}$.

Using our definition of $\varepsilon_\perp$ and applying the Cauchy-Schwarz inequality \cite{watrouscs}, we get

\begin{equation}
\| P \Delta K P\boldsymbol{{z}} \|_2 \leq \varepsilon_\perp \|\boldsymbol{z}\|_2
\end{equation}
and similarly
\begin{equation}
    \| P \Delta K P\boldsymbol{\tilde{z}} \|_2  \leq \varepsilon_\perp \|\boldsymbol{\tilde{z}}\|_2
\end{equation}

As discussed in Appendix \ref{appendixAlphaBound}, since $Y$ is a diagonal matrix with entries $\pm 1$, it preserves Euclidean norms. Therefore, we have 

$$
\|\boldsymbol{{z}}\|_2 = \|\boldsymbol{{\alpha}}\|_2 \text{ and }
\|\boldsymbol{\tilde{z}}\|_2 = \|\boldsymbol{\tilde{\alpha}}\|_2
$$

\noindent where we recall that $\boldsymbol{{\alpha}}$ is the dual solution belonging to the feasible set $\mathcal{A} = \{ \boldsymbol{\alpha} \in \mathbb{R}^m: 0 \leq \alpha_i \leq C, y^T \boldsymbol{\alpha} = 0 \}$, where all variables are defined identically as in Appendix \ref{appendixAlphaBound}.

We can determine a more conservative bound on $\|\boldsymbol{\alpha}\|_2$ using the constraint $0 \leq \alpha_i\leq C$ for some $i \in \{1,m\}$ for $m$ training samples. Using this constraint and the definition of the Euclidean norm, $\|x\|^2_2 = \sum_{i=1}^m x_i^2$ \cite{watrouscs}, we can derive the result 

\begin{equation}
\|\boldsymbol{\alpha} \|^2_2 = \sum_{i=1}^m\alpha^2_i\leq 
\sum_{i=1}^m C^2 =mC^2
\end{equation}

A similar result can be found for $\| \boldsymbol{\tilde{\alpha}} \|_2 $.

Using the above relationship and inequality (\ref{bound2})from Appendix \ref{appendixAlphaBound}, we arrive at the result,

\begin{equation}\label{worstCaseAlpha}
\| \boldsymbol{\tilde{\alpha}} - \boldsymbol{\alpha} \|_2 \leq \frac{C \sqrt{m}}{\tau}\varepsilon_\perp
\end{equation}

We can now use inequality (\ref{worstCaseAlpha}) in our stability bounds defined in the main text, which has the form 

$$
|q_\times - q_0| \leq \|\tilde{\boldsymbol{\alpha}}-\boldsymbol{\alpha}\|_2 \| Q (\boldsymbol{\tilde{\alpha}} + \boldsymbol{\alpha})\|_2
$$

\noindent where $q_\times$ and $q_0$ are the squared-inverse of the cross-kernel and ideal margins defined as $q_0 = \boldsymbol{\alpha}^TQ\boldsymbol{\alpha}$ and $q_\times = \boldsymbol{\tilde{\alpha}}^TQ\boldsymbol{\tilde{\alpha}}$.

We first consider the expression obtained in equation (\ref{qdiff}) from the main text,

$$
q_\times - q_0 = (\tilde{\boldsymbol{\alpha}}-\boldsymbol{\alpha})^T Q (\boldsymbol{\tilde{\alpha}} + \boldsymbol{\alpha})
$$. 

If we let $\boldsymbol{d} :=\tilde{\boldsymbol{\alpha}}-\boldsymbol{\alpha}$ and $\boldsymbol{s} :=\tilde{\boldsymbol{\alpha}}+\boldsymbol{\alpha}$, this expression becomes

$$ \boldsymbol{d}^T Q \boldsymbol{s} = (Y\boldsymbol{d})^T K (Y\boldsymbol{s})$$ 

\noindent since $Q = YKY$. Since both $\boldsymbol{\tilde{\alpha}}$ and $\boldsymbol{\alpha}$ are feasible, using a similar argument as in Appendix \ref{appendixAlphaBound}, we have

$$ \boldsymbol{d}^T Q \boldsymbol{s} = (Y\boldsymbol{d})^T PKP (Y\boldsymbol{s})$$

Since $Y$ preserves Euclidean norms, we get

$$ |q_\times - q_0| \leq \|\boldsymbol{d}\|_2 \|PKP\|_2 \|\boldsymbol{s} \|_2$$

Defining $\kappa := \|PKP\|_2$ yields,

$$ |q_\times - q_0| \leq \kappa \|\boldsymbol{d}\|_2  \|\boldsymbol{s} \|_2$$

Using the triangle inequality \cite{watrouscs} to get 
$$ \|\boldsymbol{s}\|_2 =\| \boldsymbol{\tilde{\alpha}} + \boldsymbol{\alpha}\|_2 \leq \| \boldsymbol{\tilde{\alpha}}\|_2 + \|\boldsymbol{\alpha}\|_2 \leq 2C\sqrt{m}
$$

\noindent Combining these results, we finally arrive at the bound

\begin{equation}\label{worstCasePBfinal}
|q_\times - q_0| \leq \frac{2mC^2 \kappa}{\tau}\varepsilon_\perp
\end{equation}

This bound is more conservative than the bound derived in the main text and used in the numerical experiments since it uses the spectral norm term, $\varepsilon_\perp$. Therefore, it is too loose to consider numerically. However, it can still be considered theoretically as a conservative stability bound for the considered quantities.

\section{Dataset Preparation}\label{AppendixE}
In this section, we provide more details about the initial preparation of the datasets used for the numerical simulations.

As mentioned in the main text, the Heart Disease, (White) Wine, HTRU2, and Breast Cancer datasets were obtained using the UCI Machine Learning Repository \cite{ucimlrepo}.  

\textbf{(White) Wine Dataset}
This dataset includes white vinho verde wine samples, with the target variable containing values quantifying wine quality. Wine quality contained multiple classes, with values ranging from $3$ to $9$.

We categorise all samples with wine quality scores of above $5$ as premium quality wine, and the remaining samples as standard quality wine. We then defined all premium quality wine to form the binary class $1$, with standard quality wine forming class $0$, thereby transforming this dataset for a binary classification task.

\textbf{Heart Disease Dataset}
The heart disease dataset consists of $13$ features and $1$ target variable. Similar to the Wine Quality dataset, this target variable contained multiple values, ranging from $0$ to $4$, where the values $\in [1,4]$, indicated the presence of heart disease in the particular patient, with $0$ indicating the absence of heart disease. The target variable has similarly been transformed for binary classification by defining all samples with heart diseases scores above $0$ as class $1$, with samples with scores of $0$ used to form class $0$. This allows the dataset to be separated into samples of patients who have heart disease, and those who do not.
Additionally, this dataset initially had $6$ missing values; $2$ and $4$ values from two features, respectively. These missing values were imputed with the median value from each feature vector using \textit{sklearn's SimpleImputer}.

\textbf{Breast Cancer Dataset}
This dataset contains $9$ features and $1$ binary target variable with values "M" or "B" indicating if a tumour is malignant or benign. This column was also transformed for binary classification by mapping malignant and benign tumours to the values $1$ and $0$, respectively. 

\textbf{HTRU2 Dataset}
The HTRU2 dataset contains pulsar candidates from the HTRU2 (South) survey. This dataset consists of $7$ feature vectors and $1$ binary target variable. This target variable takes the value of $1$ for pulsars and $0$ for non-pulsars. 

\textbf{Gaussian Dataset} 
The Gaussian dataset was curated using scikit-learn's \textit{make\_blobs} function. This function was used to generate two isotropic Gaussian blobs representing the two classes for binary classification. The distribution was generated with a cluster standard deviation of $3$.  Larger values for the cluster standard deviation result in a higher degree of overlap between the two classes resulting in non-linearly separable data.

For consistency with the SVM dual formulation used throughout this work, all binary target labels were mapped from $\{0,1\}$ to $\{-1,1\}$ prior to training the QSVMs and computing the margin-based quantities. This transformation preserves the binary classification task while matching the conventional SVM formulation used in the theoretical derivations.

\section{Description of Experimental Parameters}\label{app: ExpParam}

This section presents the experimental parameters used to conduct the numerical experiments described in Section \ref{sec: NumericalExp} in the main text. Table \ref{tab:DatasetSelec} describes the parameters used in the numerical simulations conducted to test the stability bounds under local depolarising noise. Similarly, Table \ref{tab:IBM_expParam} presents the parameters necessary to test the bounds using the real IBM device and the \textit{FakeFez} simulator. 

Both tables represent the number of qubits, unitary layers and training samples and the regularisation parameter, $C$ for the corresponding datasets. In addition to this, the particular subsets of the corresponding datasets selected for the numerical experiments on the real hardware and the \textit{FakeFez} simulator are depicted in Table \ref{tab:IBM_expParam}.

\begin{table}[h!]
    \caption{The number of samples in the training set, $m$, qubits $N$ and unitary layers, $L$ used to test the stability bounds are presented. The regularisation parameter, $C$ is also reported.}
    \label{tab:DatasetSelec}
    \begin{tabular}{>{\centering\arraybackslash}p{2cm} >{\centering\arraybackslash}p{1cm} >{\centering\arraybackslash}p{1cm} >{\centering\arraybackslash}p{1cm} >{\centering\arraybackslash}p{1cm}}
    \toprule % <-- Toprule here
    \textbf{Dataset} & \textbf{$N$} & \textbf{$L$} & \textbf{$C$} & \textbf{$m$} \\
      \midrule % <-- Midrule here
    Heart & 2 & 1 & 1 & 227\\  
      \midrule    
      Gaussian & 2 & 1 &1 & 375  \\  
        \midrule
       Breast Cancer  & 2 & 1 &1 & 455 \\
       \midrule
        Wine & 2 & 1  & 1 & 280   \\  
        \midrule
       Wine & 2 & 2  &1 & 280   \\ 
      \bottomrule % <-- Bottomrule here
    \end{tabular}
\end{table}

\begin{table}[h!]
    \caption{The number of samples in the training set, $m$, qubits $N$, unitary layers, $L$ and regularisation parameter, $C$ are presented used to test the stability bounds with the ibm{\_}fez device and the \textit{FakeFez} simulator.}
    \label{tab:IBM_expParam}
    \begin{tabular}{>{\centering\arraybackslash}p{1.5cm} >{\centering\arraybackslash}p{1.5cm} >{\centering\arraybackslash}p{0.75cm} >
    {\centering\arraybackslash}p{0.75cm} >{\centering\arraybackslash}p{0.75cm} >
    {\centering\arraybackslash}p{0.75cm}}
    \toprule % <-- Toprule here
    \textbf{Dataset} & \textbf{Samples} & \textbf{$N$} & \textbf{$L$} & \textbf{$C$} & \textbf{$m$} \\
      \midrule % <-- Midrule here
    Breast Cancer & 200-220 & 2 & 1 & 1 & 12  \\  
      \midrule    
      Breast Cancer & 0-45 & 2 & 1 &  1& 28 \\  
        \midrule
       Gaussian  & 267-302 & 2 & 1 & 1 & 20\\
      \bottomrule % <-- Bottomrule here
    \end{tabular}
\end{table}

\section{IBM Device Calibration Data}

Since the numerical experiments using the ibm{\_}fez device presented in Section \ref{ValidBounds} were obtained at two different points in time, we present two tables (Tables \ref{tab:Calib_BC} and \ref{tab:Calib_BCGaus}) containing calibration data for both runs. 

Table \ref{tab:Calib_BC} contains the calibration data for the experiment involving the subset of $20$ samples from the Breast Cancer dataset. Similarly, Table \ref{tab:Calib_BCGaus} contains the calibration data for the numerical experiments involving the subset of the Gaussian dataset (with $35$ samples), and the second subset of the Breast Cancer dataset using $45$ samples. The experiments for these two particular subsets were run consecutively and so share the same calibration data.

Both tables present the average values of the T$_1$ and T$_2$ times (in $\mu s$), and the RX,Pauli-X, CZ and RZZ errors for $156$ qubits associated with the ibm$\_$fez device. The median, range and standard deviation of these noise parameters are also included in each table. A detailed description of each error can be found in \cite{IBMQuantumDocumentationErrors}. More details regarding the calibration data for the hardware experiments can be found in \cite{Govender_The_Cross-Kernel_Margin}.

\begin{table*}[h!]
\caption{Calibration Data for the ibm{\_}fez device associated with the numerical experiment using a subset of the Breast Cancer dataset with $20$ samples. The T$_1$ and T$_2$ times were retrieved, along with single- and two-qubit gate errors for all $156$ qubits used for ibm$\_$fez. The mean, median, standard deviation, minimum and maximum of these noise parameters are presented.}
    \label{tab:Calib_BC}
    \begin{tabular}{>{\centering\arraybackslash}p{2.5cm} >{\centering\arraybackslash}p{2.5
    cm} >{\centering\arraybackslash}p{2.5cm} >{\centering\arraybackslash}p{2.5cm} >{\centering\arraybackslash}p{2.5cm} >{\centering\arraybackslash}p{2.5cm}}
    \toprule % <-- Toprule here
    \textbf{Parameter (Unit)} &
     \textbf{Average} & \textbf{Standard Deviation} & \textbf{Median} & \textbf{Minimum} & \textbf{Maximum} \\
      \midrule % <-- Midrule here
     T$_1(\mu s)$ & 111.18 & 47.07 & 87.71 & 72.08 & 198.63 \\  
      \midrule    
     T$_2 (\mu s)$ & 122.43 & 74.28 & 111.41 & 44.46 & 255.88  \\  
        \midrule
        RX Error  & 0.00037 & 0.00045 & 
0.00027 & 0.00013 & 0.0045 \\
       \midrule
        Pauli-X Error  & 0.00037 & 0.00045 & 
0.00027 & 0.00013 & 0.0045\\  
        \midrule
       CZ Error  &  0.0044 & 0.0073 & 0.0026 & 0.0014 & 0.067\\
       \midrule
       RZZ Error & 0.0079 & 0.025 & 0.0026 & 0.0012 & 0.25\\
      \bottomrule % <-- Bottomrule here
    \end{tabular}
\end{table*}

\begin{table*}[h!]
    \caption{Calibration Data for the ibm{\_}fez device associated with the numerical experiments using a subset of the Breast Cancer dataset with $45$ samples, and the Gaussian dataset using $35$ samples. The T$_1$ and T$_2$ times were retrieved, along with single- and two-qubit gate errors for all $156$ qubits used for ibm$\_$fez. The mean, median, standard deviation, minimum and maximum of these noise parameters are presented.}
    \label{tab:Calib_BCGaus}
    \begin{tabular}{>{\centering\arraybackslash}p{2.5cm} >{\centering\arraybackslash}p{2.5cm} >{\centering\arraybackslash}p{2.5cm} >{\centering\arraybackslash}p{2.5cm} >{\centering\arraybackslash}p{2.5cm} >{\centering\arraybackslash}p{2.5cm}}
    \toprule % <-- Toprule here
    \textbf{Parameter (Unit)} &
     \textbf{Average} & \textbf{Standard Deviation} & \textbf{Median} & \textbf{Minimum} & \textbf{Maximum} \\
      \midrule % <-- Midrule here
     T$_1(\mu s)$ & 
143.73 &
57.05 &
142.77 &
87.18 & 
243.35 \\
  
      \midrule    
     T$_2(\mu s)$ & 92.28 &
56.27 &
93.54 &
5.60 &
206.42 \\
  
        \midrule
        RX Error  & 0.00041 &
0.00059 &
0.00031 &
0.00013 &
0.0069 \\
 \midrule
        Pauli-X Error  & 0.00041 &
0.00059 &
0.00031 &
0.00013 &
0.0069 \\
 \midrule
       CZ Error  &  0.0048 &
0.0073 &
0.0025 &
0.0014 &
0.054 \\
   \midrule
       RZZ Error & 0.0047 &
0.0071 &
0.0026 &
0.0014 &
0.054 \\

      \bottomrule % <-- Bottomrule here
    \end{tabular}
\end{table*}

\end{document}